\documentstyle[amsfonts,12pt,oldlfont]{article}

\title{Random matrices, Virasoro algebras and\\non-commutative KP }

\author{M. Adler\thanks{Department of Mathematics,
Brandeis University, Waltham, Mass 02254, USA. E-mail:
adler@math.brandeis.edu. The support of a National Science
Foundation grant \# DMS-9503246 is gratefully
acknowledged.}~~~~~~T. Shiota\thanks{Department of Mathematics,
Kyoto University, Kyoto 606, Japan. The hospitality of the
University of Louvain and Brandeis University is gratefully
acknowledged.}~~~~~~P. van Moerbeke\thanks{Department of
Mathematics, Universit\'e de Louvain, 1348 Louvain-la-Neuve,
Belgium and Brandeis University, Waltham, Mass 02254, USA. E-mail:
vanmoerbeke@geom.ucl.ac.be and @math.brandeis.edu. The  support of
a National Science Foundation grant \# DMS-9503246, a Nato, a FNRS
and a Francqui Foundation grant is gratefully acknowledged.}}

\date{March 97}
\newcommand{\MAT}[1]{\left(\begin{array}{*#1c}}
\newcommand{\mat}{\end{array}\right)}

\newcommand{\rg}{\rightarrow}

\newcommand{\AR}{{\cal A}}
\newcommand{\BZ}{{\Bbb Z}}
\newcommand{\BC}{{\Bbb C}}
\newcommand{\Sg}{\Sigma}
\newcommand{\iy}{\infty}
\newcommand{\pl}{\partial}
\newcommand{\al}{\alpha}
\newcommand{\be}{\beta}
\newcommand{\om}{\omega}

\newcommand{\SR}{{\cal S}}
\newcommand{\PR}{{\cal P}}
\newcommand{\vp}{\varphi}

\newcommand{\dt}{\delta}
\newcommand{\Dt}{\Delta}
\newcommand{\vr}{\varepsilon}
\newcommand{\sg}{\sigma}
\newcommand{\BR}{{\Bbb R}}
\newcommand{\lb}{\lambda}

\def\BY{{\Bbb Y}}

\def\t{\underline t}
\def\Wronskian{\mathop{\rm Wronskian}}
\def\Tr{\mathop{\rm Tr}}
\def\Re{\mathop{\rm Re}}
\def\Im{\mathop{\rm Im}}
\def\vol{\mathop{\rm vol}}

\def\span{\mathop{\rm span}}

\def\be{\begin{equation}}
\def\ee{\end{equation}}
\def\bea{\begin{eqnarray}}
\def\eea{\end{eqnarray}}

\ifx\undefined\Bbb
        \let\Bbb\bf
\fi
\ifx\undefined\curvearrowright
        \let\cv\to
\else
        \newcommand{\cv}{\curvearrowright}
\fi

\catcode`\@=11
\def\ps@X{\let\@mkboth\@gobbletwo
        \def\@oddhead{\tt Adler-Van Moerbeke:%
        Random matrix\hfil\today\ \#1\hfil\S\thesection, p.\thepage
        }
        \def\@oddfoot{\rm\hfil\thepage\hfil}
        \let\@evenhead\@oddhead
        \let\@evenfoot\@oddfoot}
\catcode`@=12
\begin{document}
\maketitle

\noindent Table of contents\footnote{A corrected version
has appeared in the "Duke Mathematical journal", {\bf
94},~379--431, 1998}:
\medbreak

\noindent 1. Random matrices reviewed
\newline \noindent 2. KP revisited
\newline \noindent 3. Vertex operators for the KP equation
\newline \noindent 4. Virasoro equations satisfied by vertex operators and
$\tau$-functions
\newline \noindent 5. Fay  identities and symmetries
\newline \noindent 6. Determinant of kernels and vertex operators
\newline \noindent 7. Fredholm determinants and proof of Theorem 0.1
\newline \noindent 8. Virasoro and KP: a non-commutative KP-hierarchy
\newline \noindent 9. Proof of Theorem 0.2
\newline \noindent 10. Examples

\medbreak

\noindent {\em What is the connection of random matrices with
integrable systems}~? {\em Is this connection really useful}~?
The answer to these questions leads to a new and
unifying approach to the  theory of random matrices. Introducing an
appropriate time
$t$-dependence in the probability distribution of
the matrix ensemble, leads
to vertex operator expressions for the $n$-point
correlation functions
(probabilities of $n$ eigenvalues in infinitesimal
intervals) and the corresponding
Fredholm determinants (probabilities of no eigenvalue in a Borel subset
$E$);   the latter probability is a ratio of $\tau$-functions for the
KP-equation, whose numerator satisfy partial differential equations, which
{\em decouple} into  the sum of two parts: a Virasoro-like part depending on
time  only and a Vect($S^1$)-part depending on the boundary points $A_i$ of
$E$. Upon setting $t=0$, and using the KP-hierarchy to eliminate
$t$-derivatives, these PDE's lead to a hierarchy of non-linear PDE's, purely in
terms of the $A_i$. These PDE's are nothing else but the KP hierarchy
for which the $t$-partials, viewed as commuting operators, are replaced by
non-commuting operators in the endpoints $A_i$ of the $E$
under consideration. When the boundary of $E$ consists of one
point and for the known kernels, one recovers the Painlev\'e equations, found
in [14,24]; from
the present work, it also appears that some of the
Painlev\'e equations can be viewed as the KP equation in
non-commutative operators. Some of these results have been announced by us in
[5].

Random matrices provide a model for excitation spectra of
heavy nuclei at high excitations (Wigner [27],
Dyson [12,13] and Mehta [18,19]),
based on the nuclear experimental data by Porter and
Rosenzweig [21]; they observed that the occurrence of two levels,
close to each other, is a rare event (level repulsion),
showing that the spacing is not Poissonian, as one might
expect from a naive point of view; this lead Wigner to his
so-called surmise.

In their pioneering work, Jimbo, Miwa, Mori and SatoÊ[14] have
shown some $(p,q)$-variables derived from the sine kernel satisfy a
certain Neumann-like completely integrable finite-dimensional
Hamiltonian system. They also showed that the distribution of the
level-spacings for the sine kernel satisfies a Painlev\'e equation. Tracy
and Widom [24,25] have successfully used functional-theoretical tools to
compute the level spacing distributions for a more general class of kernels,
always yielding Neumann-type systems. Deift, Its and Zhou [10] have used
the Riemann-Hilbert approach to find the precise asymptotics for the
distributions above. Random matrices have come up in the context of statistical
mechanics and quantum gravity; see [7, 15, 17].
 Random matrix ideas play an
increasingly prominent role in mathematics:
not only have they come up in the spacings of
the zeroes of the Riemann zeta function, but their relevance
has been observed in the chaotic Sinai billiard and, more
generally, in chaotic geodesic flows; Sarnak\cite{S}
conjectures that chaos leads to the ``spectral
rigidity", typical of  the spectral distributions of random
matrices,  whereas the spectrum of an integrable system is
random (Poisson)!

As is well known in the theory of random matrices, probability
distributions of finite matrix ensembles are usually given by explicit matrix
integrals, whereas the probabilility measure for infinite matrix ensembles is
given by certain kernels. We found out that the first situation,
actually leading to Christoffel-Darboux kernels, typically ties up with the
Toda lattice and the second, corresponding to kernels represented
by an integral, ties up with the KdV equation. This paper deals with the second
case (KdV) only. There is a striking analogy between the two cases, showing our
point of view is very robust. A similar theory can be worked out for
finite matrix ensembles (discrete kernels and 1-Toda lattice) and for two
coupled random matrices (2-Toda lattice). Theorems 0.1 and 0.2, stated
in this section, are the main results of this paper.

In this work, we shall be dealing with the $KP$ hierarchy, briefly explained in
section 2; it is a hierarchy of isospectral deformations of a
pseudo-differential operator $L = D +\sum_{i \geq 1} a_i (x,t) D^{-i}$,
with\footnote{$A_+$ denotes the differential part of the pseudo-differential
operator $A$}
$D:=d/dx$,
\be
\frac{\pl L}{\pl t_n } = [(L^n )_+,L ]~, \mbox{ for }~~ t \in \Bbb{C}^{\infty}.
\ee
We also consider the $p$-reduced $KP$ hierarchy, i.e. the reduction to $L$'s
such that
$L^p $ is a differential operator for some fixed $p \geq 2$.  Sato tells us that
the solution $L$ to equations (1) can ultimately be expressed in terms of a
$\tau$-function.
The wave and adjoint wave functions, expressed in terms of the
$\tau$-function\footnote
{$[\al]=\left(\al,\frac{\al^2}{2},\frac{\al^3}{3},...\right)$}
\be
\Psi (x,t,z) = e^{xz + \sum_1^{\infty} t_i z^i } \frac{\tau (t-[z^{-1} ])}{\tau
(t)} ~~\mbox{and}~~\Psi^* (x,t,z) = e^{-xz - \sum_1^{\infty} t_i z^i }
\frac{\tau (t + [ z^{-1}])} {\tau (t)} ;
\ee
 satisfy
\be
\begin{array}{ll}
z \Psi = L \Psi &\quad~~~ z \Psi^* = L^{\top} \Psi^* \\
\frac{\partial \Psi}{\partial t_n } = (L^n )_+ \Psi &\quad ~~~\frac{\partial
\Psi^*}{\partial t_n } = - (L^{\top n} )_+ \Psi^* .
\end{array}
\ee

In this paper, as in the general theory of integrable systems, vertex
operators play a prominent role : they are Darboux transforms involving all
times.  In particular, for the $KP$ equation, the vertex operator
\be
X(t,y,z) := \frac{1}{z-y} e^{\sum_1^{\infty} (z^i-y^i )t_i}
e^{\sum_1^{\infty} (y^{-i}-z^{-i} ) \frac{1}{i} \frac{\partial}{\partial
t_i}}
\ee
has the striking feature that $X(t,y,z) \tau$ and $\tau + X (t,y,z) \tau$ are
both $\tau$-functions. Given distinct roots of unity
$\omega,\omega' \in \zeta_p := \{
\om ~|~\om^p = 1\}$, the vertex operator $X(t,\omega z,\omega' z)$ maps the
space of p-reduced $\tau$-functions into itself; for vertex operators and their
interaction with the Virasoro algebra, see sections 3 and 4.

Sections 3,4, 5 and 6 will lead to the first main theorem, the first part of
which is a continuous version of a ``soliton" formula. These sections and
Theorem 0.1 are due to Adler-Shiota-van Moerbeke \cite{ASV3}. For the
sake of completeness, we sketch the proof in this paper.

\medbreak

\proclaim Theorem 0.1. Define the $(x,t)$-dependent kernel $k_{x,t} (y,z)$ and
$k^E_{x,t}(y,z):= k_{x,t} (y,z) I_E (z)$ with $x \in \BR, ~t \in \BC^{\infty}$,
$y,z \in \BC$, and $E \subset \Bbb{R}^+ $ a Borel subset:
\be
k_{x,t} (y,z): = \int^x dx \sum_{\omega \in \zeta_p } a_{\omega} \Psi^*
(x,t,\omega y) \sum_{\omega' \in \zeta_p } b_{\omega'} \Psi (x,t,\omega' z),
\ee
where $\Psi (x,t,z)$ and $\Psi^* (x,t,z)$ are the wave and adjoint wave
function for the $p$-reduction of $KP$ and where the coefficients $a_{\omega}
,b_{\omega} \in \Bbb{C}$ are subjected to $\sum_{\omega \in \zeta_p }
\frac{a_{\omega} b_{\omega} }{\omega} = 0$.  Then the following holds:
\medbreak
\noindent
(i) the kernel $k(y,z)$, its determinant and its
Fredholm determinant are all three expressible in terms of the vertex
operator
\be
Y(t,y,z):=\sum_{\om,\om'\in\zeta_p}a_{\om}b_{\om'}X(t,\om
y,\om'z)
\ee
acting on the underlying $\tau$-function\footnote{$Y(t,z,z)$ in
formula (7) below must be understood as $\lim_{y\rg z}Y(t,y,z)$.} :
$$k_{x,t}(y,z) = \frac{1}{\tau}  Y(t,y,z) \tau $$
$$\det (k_{x,t} (y_i ,z_j ))_{1Ê\leq i,j \leq n} = \frac{1}{\tau} Y(t,y_1 ,z_1
) \ldots Y(t,y_n ,z_n ) \tau $$
\be
\det (I -Ê\mu k^{E}_{x,t} ) = \frac{1}{\tau} e^{-Ê\mu \int_E dz~
Y(t,z,z)} \tau
\ee
$$\quad\quad\quad\quad\quad\quad\quad\quad\quad\quad\quad\quad
\quad\quad\quad\quad\quad\mbox{(``continous" soliton formula)}
$$
\newline
\noindent(ii) Let the kernel $k_{x,t}
(y,z)$ in (5) be such that the underlying $\tau$-function of $\Psi$ and
$\Psi^*$ satisfies a
Virasoro constraint~\footnote{Define $W_n^{(0)}=\delta_{n,0}$,
$$
J_n^{(1)}:=W_n^{(1)}=\left\{\begin{array}{ll}
\pl/\pl t_n     & \hbox{if }n>0\\
(-n)t_{-n}      & \hbox{if }n<0\\
0               & \hbox{if }n=0
\end{array}\right.
,\quad
J_n^{(2)}:=W_n^{(2)}+(n+1)W_n^{(1)}
=\sum_{i+j=n}\mathopen:J_i^{(1)}J_j^{(1)}\mathclose:
$$}:
$$W^{(2)}_{kp} \tau = c_{kp} \tau \quad\mbox{ for a fixed } k \geq -1.$$
Then for the disjoint union
$E =\bigcup^r_{i=1} [a_{2i-1} ,a_{2i} ] \subset \Bbb{R}_+$,
the Fredholm determinant $\det (I-\mu  k^{E}_{x,t})$
satisfies the following Virasoro constraint for that same $k \geq -1$:
\be
\left(-\sum_{i=1}^{2Êr} a^{kp+1}_i \frac{\partial}{\partial a_i } +
\frac{1}{2} (W^{(2)}_{kp} - c_{kp} )\right) (\tau \det (I - \mu k^E_{x,t} )) =
0;
\ee
note the {\em boundary} $a$-part and the {\em time} $t$-part decouple. \par
In the next theorem we apply equation (8) to compute the partial differential
equations for the distribution of the spectrum for matrix ensembles whose
probability is given by a kernel.  To state the problem, consider a first-order
differential operator $A$ in $z$ of the form
\be
A=A_z=\frac{1}{2}z^{-m+1}\left(\frac{\pl}{\pl
z}+V'(z)\right)+\sum_{i\geq 1}c_{-2i}z^{-2i},
\ee
with
\be
V(z)=\frac{\al}{2}z+\frac{\beta}{6}z^3\not\equiv 0,\quad
m=\deg V'=0\mbox{\,\,or\,\,}2,
\ee
and the differential part of its ``Fourier" transform
\be
\hat A=\hat
A_x=\left(\frac{1}{2}\left(x+V'(D)\right)D^{-m+1}+\sum_{i\geq
1}c_{-2i}D^{-2i}\right)_+\quad \mbox{with} \quad D=\frac{\pl}{\pl x}.
\ee
Given a disjoint union
$E=\displaystyle{\bigcup^r_{i=1}}[A_{2i-1},A_{2i}] \subset \BR^+$, define
differential operators $\AR_n$, which we declare to be of homogeneous ``weight"
$n$, as follows
$$
\AR_n:=\sum_{i=1}^{2r}A_i^{\frac{n+1-m}{2}}\frac{\pl}{\pl A_i},
\quad n=1,3,5,... .
$$
We now state the second main theorem, established in section 8:

\proclaim Theorem 0.2.  Let $\Psi(x,z)$, $x\in\BR$, $z\in\BC$ be a solution
of the linear partial differential equation
\be
A_z\Psi(x,z)=\hat A_x\Psi(x,z),
\ee
with holomorphic (in $z^{-1}$) initial condition at $x=0$,
subjected to the following differential equation for some $a,b,c
\in \BC$,
\be
(aA^2_z+bA_z+c)\Psi(0,z)=z^2\Psi(0,z),\mbox{\,\,with\,\,}
\Psi(0,z)=1+O(z^{-1}).
\ee
Then\newline\indent $\bullet~~~\Psi(x,z)$ is a
solution of a second order problem for some potential $q(x)$
\be
(D^2+q(x))\Psi(x,z)=z^2\Psi(x,z).
\ee
\indent $\bullet~~~$ Given the
kernel
\be
K^E_x(y,z):=I_E(z)\int^x
\frac{\Phi(x,\sqrt{y})\Phi(x,\sqrt{z})}{2y^{1/4}z^{1/4}}
dx,
\ee
with
$$
\Phi(x,u):=\sum_{\om=\pm 1}b_{\om}e^{\om V(u)}\Psi(x,\om
u),
$$
the Fredholm determinant $f(A_1,...,A_{2r}):=\det(I-\lb
K_x^E)$ satisfies a hierarchy of bilinear partial
differential equations
\footnote{the
$p_i$ are the elementary Schur polynomials
$e^{\sum_1^{\iy}t_iz^i}=\sum_0^{\iy}p_n(t)z^n$, and $p_i(\pm\tilde{\cal
A}):=p_i\displaystyle{(\pm{\cal A}_1,0,\pm\frac{1}{3}{\cal
A}_3,0,...)}$}
in the $A_i$ for odd $n \geq 3$:
$$
f \cdot \AR_n \AR_1 f-\AR_n f \cdot \AR_1 f-\sum_{i+j=n+1}p_i
(\tilde\AR)f \cdot p_j(-\tilde\AR)f
$$
\be
+ \,(\mbox{terms of lower weight $i$ for $1\leq i\leq n)=0$},
\ee
where $x$ appears in the coefficients of the lower weight terms only.

\bigbreak

\section{Random matrices
reviewed}

Define on the ensemble ${\cal H}_N=\{N\times N\hbox{ Hermitian matrices}\}$
the probabi\-li\-ty~~\footnote{
        $dM=\prod_1^NdM_{ii}\prod_{1\le i<j\le N}d\Re(M_{ij})d\Im(M_{ij})$
}
$$
P(M\in dM)=ce^{-\Tr V(M)}dM,
$$
where $c$ is a normalization constant.  Then for
$z_1,\dots,z_N\in\BR$ we have~~\footnote{
        $\Delta(z)=\prod_{1\le i<j\le N}(z_i-z_j)$,
the Vandermonde determinant
}
\begin{eqnarray*}
\lefteqn{
P(\hbox{one eigenvalue in each $[z_i,z_i+dz_i]$, $i=1,\dots,N$})
}\\
&=&c\vol(U(N)) e^{-\sum_1^N V(z_i)}\Delta^2(z) dz_1\cdots dz_N
\end{eqnarray*}
and for $0\le k\le N$
\begin{eqnarray}
\lefteqn{
P(\hbox{one eigenvalue in each $[z_i,z_i+dz_i]$, $i=1,\dots,k$})
}\nonumber\\
&=& c'\left(\vphantom{\int\limits_a}
\smash{\mathop{\int\!\cdots\!\int}_{\BR^{N-k}}}
e^{-\sum_1^N V(z_i)}\Delta^2(z) dz_{k+1}\cdots dz_N\right)dz_1\cdots dz_k
\nonumber\\
&{\buildrel*\over=}&
c''\det\left(K_N(z_i,z_j)\right)_{1\le i,j\le k} dz_1\cdots dz_k,
\label{15}
\end{eqnarray}
and if $J\subset \BR$, then
$$
P(\hbox{exactly $k$ eigenvalues in $J$})=
{(-1)^k\over k!}
{\pl^k\over\pl\lb^k}\det\left.\left(I-\lb K_N^J\right)\right|_{\lb=1},
$$
where
$$
K_N^J(z,z')=K_N(z,z')I_J(z'),
$$
$I_J$ the indicator function of $J$, and $K_N$ is the Schwartz kernel of
the orthogonal projector $\BC[z]\to\BC+\BC z+\cdots+\BC z^{N-1}$ with
respect to the measure $e^{-\Tr V(z)}dz$, namely
$$
K_N(z,z')=\sum_{k=0}^{N-1}\vp_k(z)\vp_k(z')
$$
in terms of orthonormal functions $\vp_k(z)=e^{-\Tr V(z)/2}p_k(z)$
with respect to $dz$ or orthogonal polynomials
$p_k(z)=(1/\sqrt{h_k})z^k+\cdots$ with respect to $e^{-\Tr V(z)}dz$.

In the last equality (${\buildrel*\over=}$) in (15) one has used
the simple property of a Vandermonde that
\begin{eqnarray*}
\Delta(x)^2&=&\det\left(q_{k-1}(x_i)\right)_{1\le i,k\le N}^2\\
&=&\det\left(\sum_{k=1}^N q_{k-1}(x_i)q_{k-1}(x_j)\right)_{1\le i,j\le N}
\end{eqnarray*}
in terms of any monic polynomials $q_k(x)$ of degree $k$. Then in
(${\buildrel*\over=}$) the orthonormality of $\vp_k$ is used to obtain
$$
        \int K_N(x,y)K_N(y,z)dy = K_N(x,z),
$$
which just means that $K_N$ gives a projector. When $V(z)$ is quadratic, we
have for large $N$,
\begin{eqnarray*}
        P(\hbox{an eigenvalue}\in [z,z+dz]) &=& K_N(z,z)dz\\
        &\sim&
        \left\{\begin{array}{ll}
                {1\over\pi}(2N-z^2)^{1/2}dz & \hbox{if }|z|<(2N)^{1/2}\\
                0 & \hbox{if }|z|>(2N)^{1/2}
        \end{array}\right.
\end{eqnarray*}
is given by the circular distribution (Wigner's semi-circle law), we have that
for
$z\sim 0$ the average spacing between the eigenvalues near the origin is
$\sim (K_N(0,0))^{-1}=\pi/\sqrt{2N}$ and near the edge ($z\sim\sqrt{2N}$)
is $1/(2^{1/2}N^{1/6})$, leading to
$$
\lim_{N\uparrow\infty}{1\over K_N(0,0)}K_N\left(
        {z\over K_N(0,0)}, {z'\over K_N(0,0)}
\right)
=K(z,z')={1\over\pi}{\sin\pi(z-z')\over z-z'}
\eqno(\hbox{bulk scaling limit})
$$
$$
\lim_{N\uparrow\infty}{1\over K_N(0,0)}K_N\left(
        \sqrt{2N}+{z\over 2^{1/2}N^{1/6}}, \sqrt{2N}+{z'\over 2^{1/2}N^{1/6}}
\right)
=\int_0^\infty A(x+z)A(x+z')dx
\eqno(\hbox{edge scaling limit})
$$
in terms of the classical Airy function.  In a similar context one also finds
the Bessel kernel; for background on such matters, consult Mehta's excellent
book \cite{M1}.

\section{KP revisited}\label{Three}

Consider the infinite-dimensional span
$$
W=\mbox{span}_{\BC}\{\psi_k(z)=z^k(1+O(z^{-1})),k=0,1,2,...\}
$$
of meromorphic functions $\psi_k(z)$ in $z^{-1}$ with poles of order $k$ at
$z=\iy$. With Sato, define the $\tau$-function
\footnote{$H_+=\mbox{span}_{\BC}\{1,z,z^2,...\}$}
$$
\tau(x,t)=\tau(\t):=\det\mbox{\,Proj\,}\left(e^{-\sum_1^{\iy}
\t_iz^i}\rg
H_+\right)
$$
where $\underline{t}=t+(x,0,0,...)=(x+t_1,t_2,...)$. The $\tau$-function is
shown to satisfy the following bilinear relation,
\bea
\oint e^{\sum_1^{\iy}
(t_i-t'_i)z^i}\tau(t-[z^{-1}])\tau(t'+[z^{-1}])dz= 0,\label{bilinear}
\eea
where the integral is taken over a small circle around $z=\iy$; this relation
actually characterizes the $\tau$-function.

The tau-function $\tau(t)$ above leads a wave operator
\bea
S:=\sum^{\iy}_{n=0}\frac{p_n(-\tilde\pl)\tau(\t)}{\tau(\t)}D^{-n},
\eea
and also to a wave and dual wave function
\bea
\Psi(x,t,z):=(S\,e^{xz})e^{\sum_1^{\iy}t_iz^i}=e^{xz+
\sum_1^{\iy}t_iz^i}\frac{\tau(t-[z^{-1}])}{\tau(t)},\label{Psi1}
\eea
\bea
\Psi^*(x,t,z)=((S^{\top})^{-1}e^{-xz})e^{-\sum_1^{\iy}t_iz^i}=e^{-xz-
\sum_1^{\iy}t_iz^i}\frac{\tau(t+[z^{-1}])}{\tau(t)}.\label{Psi2}
\eea
From (18) it follows that, along a small contour $z=\iy$,
\be
\int_{z=\iy}\Psi(x,t,z)\Psi^*(x',t',z)dz=0.
\ee
Conversely, if functions $\Psi$ and $\Psi^*$, with the appropriate asymptotic
behavior, satisfy the bilinear identity (22), then $\Psi$ is a wave function
and $\Psi^*$ an adjoint wave function of the KP hierarchy; i.e., there exists a
monic first order pseudodifferential operator $L=D + \cdots=S\,D\,S^{-1}$ in
$x$, depending on
$t$, satisfying (1) and (3). For more details on these matters, see the work
of Date, Jimbo, Kashiwara, Miwa \cite{Date} and for vertex operators and
symmetries, see \cite{ASV1,vM}.

Since $x$ and $t_1$ always appear together as $x+t_1$ in $\Psi$ and $\Psi^*$,
in every formula involving $\Psi$ or $\Psi^*$ and $\tau$, we always have
$\t$ instead of $t$ in the argument of $\tau$.  To simplify the notation in
such a formula, we will often omit underlining and simply write $t$ instead of
$\t$ in the argument of $\tau$, if there is no fear of confusion.

We also introduce the Orlov-Schulman pseudo-differential operator $M$,
\bea
M:=S(x+\sum_1^{\iy}kt_k\,D^{k-1})S^{-1}
\label{M}
\eea
satisfying $[L,M]=1$, and
$$
\frac{\pl\Psi}{\pl z}=M\Psi.
$$

The $p$-reduction of KP ($p\geq 2$) is one where $L^p$ is a differential
operator; it is equivalent to the property that
$$
W=\span\left\{\Psi(0,z),\frac{\pl}{\pl x}\Psi(0,z),...\right\}
$$
satisfies $z^pW\subset W$.

\medbreak

For future use, we also introduce
the
$\delta$-function,
$$
\dt(t)=\sum^{\iy}_{n=-\iy} t^n =\frac{1}{1-t}+\frac{t^{-1}}{1-t^{-1}},
$$
with the customary property
\be
f(\lb,\mu)\dt(\frac{\lb}{\mu})=f(\lb,\lb)\dt(\frac{\lb}{\mu}).
\label{delta1}
\ee
Note the function
$$\dt(\lb,\mu):=\frac{1}{\mu}\dt(\frac{\lb}{\mu})=\frac{1}{\mu}
\sum_0^{\infty}(\frac{\lb}{\mu})^n
$$
has the usual $\dt$-function property
$$\frac{1}{2\pi i} \int f(\lb,\mu)  \dt(\lb,\mu)d\mu=f(\lb,\lb)
$$
and is a function of $\lb-\mu$ only, since
\be
\left(\frac{\pl}{\pl \lb} +\frac{\pl}{\pl \mu}\right)\dt(\lb,\mu)=0.
\label{delta2}
\ee

The action of $D^{-1}$ on ``oscillating functions'' like
$\Psi$, $\Psi^*$ is usually defined formally by $D^{-1}e^{xz}=z^{-1}e^{xz}$
and the Leibniz rule.
E.g., if $f(x,t,z)$ behaves nicely near $z=\infty$, then
\begin{eqnarray*}
        D^{-1}(f(x,t,z)e^{xz})
        &=& \sum_{k=0}^\infty\left({-1\atop k}\right)D^kf\cdot D^{-1-k}e^{xz}\\
        &=& \sum_{k=0}^\infty(-1)^kD^kf\cdot z^{-1-k}e^{xz}.
\end{eqnarray*}
When this interpretation is not sufficient, e.g., if $z$ needs to be finite
rather than defined around $\infty$, it is defined
so that it gives back this formal interpretation asymptotically as $z\to\infty$,
e.g.,
$$
        D^{-1}(f(x,t,z)e^{xz})=\left\{
        \begin{array}{rl}
                \displaystyle
                \int_{-\infty}^x f(x,t,z)e^{xz}dx & \hbox{if }\Re z>0,\\[9pt]
                \displaystyle
                -\int_x^{\infty} f(x,t,z)e^{xz}dx & \hbox{if }\Re z<0,
        \end{array}
        \right.
$$
$$
        D^{-1}(\Psi(x,t,\mu)\Psi^*(x,t,\lb))=\left\{
        \begin{array}{rl}
                \displaystyle
                \int_{-\infty}^x\Psi(x,t,\mu)\Psi^*(x,t,\lb)dx
                        & \hbox{if }\Re \mu>\Re \lb,\\[9pt]
                \displaystyle
                -\int_x^{\infty}\Psi(x,t,\mu)\Psi^*(x,t,\lb)dx
                        & \hbox{if }\Re \mu<\Re \lb,
        \end{array}
        \right.
$$
if the right hand side makes sense.  Hence we shall also denote $D^{-1}$ by
$\displaystyle\int^xdx$ when it is acting on an oscillating function.
The lower limit of this integral sign, while not written out, is important
and must depend, e.g., on the sign of $\Re z$ or $\Re\mu-\Re\lb$ in
these examples.

\section{Vertex operators for the KP equation}

Define the vertex operator as in (4):
and more generally~\footnote{where $:~:$ denotes normal
ordering, i.e., always pull differentiation
to the right.}
$$
X(t,\lb_1,\dots,\lb_n;\mu_1,\dots,\mu_n)
\quad\quad\quad\quad\quad\quad\quad\quad\quad\quad\quad\quad\quad
\quad\quad\quad\quad\quad\quad\quad\quad\quad\quad\quad\quad\quad
\quad\quad\quad\quad
$$
\begin{eqnarray}
&:=&
\frac{1}{\prod_{k\neq\ell}(\mu_k-\lb_{\ell})}\mathopen:\prod^n_1
X(\lb_i,\mu_i)\mathclose:\label{vertexcomp}\\
&:=&\frac{1}{\prod_{k,\ell}(\mu_k-\lb_{\ell})}
\prod_{k=1}^ne^{\sum_{i=1}^{\iy}(\mu^i_k-\lb^i_k)t_i}
\prod_{k=1}^ne^{\sum_{i=1}^{\iy}(\lb^{-i}_k-\mu^{-i}_k)\frac{1}{i}
\frac{\pl}{\pl t_i}},\nonumber
\end{eqnarray}
with the customary expansion of $X(t,\lb,\mu)$ in terms of
$W$-generators:
\begin{equation}
X(t,\lb,\mu)
=\frac{1}{\mu-\lb}\sum_{k=0}^{\iy}\frac{(\mu-\lb)^k}{k!}
\sum_{\ell=-\iy}^{\iy}\lb^{-\ell -k}W_{\ell}^{(k)}.
\label{Wexp}
\end{equation}
Observe that $W_{\ell}^{(0)}, W_{\ell}^{(1)}$ and $W_{\ell}^{(2)}$ coincide
with the generators in footnote 3. Vertex operators $X(t,\lb,\mu)$ are known to
satisfy the following commutation relations, due to \cite{Date}:
\begin{equation}
\left[X(\lb,\mu),X(u,v)\right]
=-\delta(u,\mu)X(\lb,v)
        +\delta(\lb,v)X(u,\mu).
\label{vertexcomm}
\end{equation}

We also define an operator $N$, which has a formal expansion similar to the
vertex operator expansion (\ref{Wexp}), but with $W^{(k)}_{\ell}$ replaced by
$z^{\ell+k-1}\left( \pl / \pl z
\right)^{k-1}$:
\begin{eqnarray}
n(\lb,\mu,z)&:=&\dt(\lb,z)e^{(\mu-\lb)\frac{\pl}{\pl z}}\nonumber \\
            &=& \sum^{\infty}_{\ell=-\infty}
\frac{z^{\ell-1}}{\lb^{\ell}}
\sum^{\infty}_{k=1}\frac{(\mu-\lb)^{k-1}}{(k-1)!}\left( \frac{\pl}{\pl z}
\right)^{k-1} \nonumber\\
                                 &=&\frac{1}{\mu-\lb} \sum_{k=0}^{\iy}\frac{(\mu-\lb)^k}{k!}
\sum_{\ell=-\iy}^{\iy}\lb^{-\ell}kz^{\ell-1}\left( \frac{\pl}{\pl z}
\right)^{k-1} \nonumber \\
             &=&\frac{1}{\mu-\lb} \sum_{k=0}^{\iy}\frac{(\mu-\lb)^k}{k!}
\sum_{\ell=-\iy}^{\iy}\lb^{-\ell -k}kz^{\ell+k-1}\left( \frac{\pl}{\pl z}
\right)^{k-1}
\label{N}
\end{eqnarray}

The next Lemma shows the precise relationship between the vertex
operator (\ref{vertexcomp}) and the composition
of vertex operators (4):

\proclaim Lemma 3.1. If, possibly after some relabeling,
$|y_1|,|z_1|<|y_2|,|z_2|<...$,  then the following relation holds:
\be
X(y_1;z_1)...X(y_n;z_n)=(-1)^{\frac{n(n-1)}{2}}
\Delta(z)\Delta(y)X(y_1,...,y_n;z_1,...,z_n).
\label{vertexrel}
\ee

\medbreak
\noindent\underline{Proof}: Whenever the inequalities above are satisfied,
the vertex operators automatically commute. Therefore it suffices to
prove (\ref{vertexrel}) for $|y_1|,|z_1|>|y_2|,|z_2|>...$; setting
$$
c(y,z):=\frac{1}{(z_1-y_1)\prod_{k,\ell=2}^n(z_k-y_{\ell})},
$$
we proceed by induction:
\begin{eqnarray*}
&&\quad X(y_1,z_1)X(y_2,...,y_n;z_2,...,z_n)\\
&=&c(y,z) e^{\sum_i(z^i_1-y^i_1)t_i}
e^{\sum_i(y_1^{-i}-z_1^{-i})\frac{1}{i}\frac{\pl}{\pl t_i}}
\quad e^{\sum^n_{k=2}\sum_i(z^i_k-y^i_k)t_i}
e^{\sum^n_{k=2}\sum_i(y^{-i}_k-z_k^{-i})\frac{1}{i}
\frac{\pl}{\pl t_i}}\\
&=&c(y,z) e^{\sum_i(z^i_1-y^i_1)t_i}e^{\sum^n_{k=2}\sum_i(z^i_k-y^i_k)
(t_i+\frac{1}{i}y_1^{-i}-\frac{1}{i}z_1^{-i})}e^{\sum^n_{k=1}
\sum_i(y^{-i}_k-z^{-i}_k)\frac{1}{i}\frac{\pl}{\pl t_i}}\\
&=&c(y,z) e^{\sum_{k=1}^n\sum^{\iy}_{i=1}\frac{1}{i}\Bigl(
(\frac{z_k}{y_1})^i+(\frac{y_k}{z_1})^i-(\frac{y_k}{y_1})^i-
(\frac{z_k}{z_1})^i\Bigr)}e^{\sum_{k=1}^n\sum^{\iy}_{i=1}
(z_k^i-y^i_k)
t_i}e^{\sum_{k=1}^n\sum^{\iy}_{i=1}(y^{-i}_k-z^{-i}_k)\frac{1}{i}
\frac{\pl}{\pl t_i}}\\
&=&c(y,z)\prod^n_{k,\ell=1}(z_k-y_{\ell})\prod^n_{k=2}\frac{(1-\frac{y_k}{y_1})(1-\frac{z_k}{z_1})}
{(1-\frac{z_k}{y_1})(1-\frac{y_k}{z_1})}X(y_1,...,y_n;z_1,...,z_n)\\
&=&(-1)^{n-1}\prod^n_{k=2}(y_1-y_k)(z_1-z_k)
X(y_1,...,y_n;z_1,...,z_n),
\end{eqnarray*}
where we have used
\be
e^{-\sum_1^{\iy}\frac{a^i}{i}}=1-a\quad\quad |a|<1~~\mbox{and\,\,}
|\frac{y_k}{y_1}|,|\frac{z_k}{z_1}|,|\frac{z_k}{y_1}|,
|\frac{y_k}{z_1}|<1;
\label{log}
\ee
this ends the proof of Lemma 3.1.

\proclaim Corollary 3.1.1. Away from $y=z$, we have
\be
X(y,z)X(y,z)=0
\label{vertexsq}
\ee
and so
\be
e^{aX(y,z)}=1+aX(y,z).
\label{vertexexp}
\ee

\medbreak
\noindent\underline{Proof}: Indeed from (\ref{vertexrel}), we have, assuming
$|y_1|,|z_1|<|y_2|,|z_2|$
$$
X(y_1;z_1)X(y_2;z_2)=-(y_1-y_2)(z_1-z_2)X(y_1,y_2;z_1,z_2)
$$
and thus, in the limit, when $y_1\longrightarrow y_2$, $z_1\longrightarrow
z_2$,  we find, away from $y_1=z_1$, the vanishing
(\ref{vertexsq}) of the square of the vertex operator.
 Thus Taylor expanding $e^{aX(y,z)}$
leads to (\ref{vertexexp}).

\proclaim Lemma 3.2. \cite{ASV3} The following commutation relation holds:
\begin{equation}
\left(v^{\ell+k}\left({\pl\over\pl v}\right)^k
        -\left(-{\pl\over\pl u}\right)^k\circ u^{\ell+k}
        \right)
        X(u,v)=
        {1\over k+1}
        \left[W_{\ell}^{(k+1)},X(u,v)\right]
        \,.
        \label{vertexvir}
\end{equation}

\medbreak
\noindent\underline{Proof}: First observe, using the property
(\ref{delta1}) of the $\delta$-function, that
\begin{eqnarray*}
n(\lb,\mu,v) X(u,v)
     &=&\dt(\lb,v)e^{(\mu-\lb)\frac{\pl}{\pl v}} X(u,v)\\
     &=& \dt(\lb,v) X(u,v+\mu-\lb)\\
     &=&\dt(\lb,v) X(u,\mu)
\end{eqnarray*}
and, using $\delta(y,z)$ is a function of $y-z$ (see
(\ref{delta2})), that

\begin{eqnarray*}
n(\lb,\mu,u)^{\top} X(u,v)
     &=&e^{(\lb-\mu)\frac{\pl}{\pl u}}\dt(\lb,u) X(u,v)\\
     &=& \dt(\lb,u+\lb-\mu) X(u+\lb-\mu,v)\\
     &=&\dt(u,\mu) X(u+\lb-\mu,v)\\
      &=&\dt(u,\mu) X(\lb,v).
\end{eqnarray*}
From the commutation relation (\ref{vertexcomm}) and the the two relations
above, it follows that
\begin{eqnarray*}
\left[X(\lb,\mu),X(u,v)\right]
&=&-\delta(u,\mu)X(\lb,v)+\delta(\lb,v)X(u,\mu)\\
&=&\left(-n(\lb,\mu,u)^{\top}+n(\lb,\mu,v)\right) X(u,v)  .
\end{eqnarray*}
Finally, expanding both sides of
$$
\left[X(\lb,\mu),X(u,v)\right]
=(n(\lb,\mu,v)-n(\lb,\mu,u)^{\top})X(u,v)
$$
in $\lb$ and $\mu$ according to (\ref{Wexp}) and (29), yields
$$
\frac{1}{\mu-\lb}\left[ \sum_{k=0}^{\iy}\frac{(\mu-\lb)^k}{k!}
\sum_{\ell=-\iy}^{\iy}\lb^{-\ell -k}W_{\ell}^{(k)},X(u,v)
\right]
$$
$$
=\sum_{k=1}^{\iy}\frac{(\mu-\lb)^{k-1}}{k!}
\sum_{\ell=-\iy}^{\iy}\lb^{-\ell -k}k \left(v^{\ell+k-1}( \frac{\pl}{\pl
v}  )^{k-1}
-(-\frac{\pl}{\pl u})^{k-1}
u^{\ell+k-1}\right)X(u,v),$$
which, upon comparing coefficients, ends the proof of Lemma 3.2.

\bigbreak

\proclaim Corollary 3.2.1. For all $k \in \Bbb Z$, we have
\be
        (v^{\ell}-u^{\ell})X(u,v)=\left[
        W_{\ell}^{(1)},X(u,v)
        \right]
\label{cor1}
\ee
and
\be
\left( v^{\ell+1}{\pl\over\pl v}+{\pl\over\pl u}u^{\ell+1}\right)
X(u,v)={1\over2}\left[W_{\ell}^{(2)},X(u,v)
\right]\,,
\label{cor2}
\ee
which further translates into the following statement for
``$(1/2,1/2)$-differentials",
$$
 (v^{\ell}-u^{\ell}) X(u,v) \sqrt{du \, dv}=
\left[W_{\ell}^{(1)}, X(u,v) \sqrt{du \, dv} \right]
$$

$$
(u^{\ell+1}\frac{\pl}{\pl u}+ v^{\ell+1}\frac{\pl}{\pl
v})X(u,v)\sqrt{du\,dv}=\left[ W^{(2)}_{\ell}+(\ell +1)W_{\ell}^{(1)},X(u,v)
\sqrt{du\,dv}\right].
$$

\medbreak
\noindent\underline{Proof}: The first two formulas follow at once from setting
$k=0$~and~1 in  (\ref{vertexvir}).
The third relation follows immediately from (\ref{cor1}). The last one
follows from combining the two computations below: on the one hand, both
(\ref{cor1}) and (\ref{cor2}) lead to
\begin{eqnarray*}
\frac{1}{2}\left[J_k^{(2)},X(u,v)\right]
&=&\frac{1}{2}\left[W_k^{(2)}+(k+1)W_k^{(1)},X(u,v)\right]\\
&=&  \left(v^{k+1}\frac{\pl}{\pl v}+\frac{\pl}{\pl u}u^{k+1}+
\frac{k+1}{2} (v^k-u^k)\right)X(u,v) \\
&=&\frac{1}{2}\Biggl(u^{k+1} \frac{\pl}{\pl u}+\frac{\pl}{\pl u} u^{k+1}
+v^{k+1} \frac{\pl}{\pl v} +
\ \frac{\pl}{\pl v} v^{k+1}\Biggr)
X(u,v);
\end{eqnarray*}
on the other hand, we have
\begin{eqnarray*}
(u^{k+1} \frac{\pl}{\pl u})(f(u)\sqrt{du})
&:=&\frac{\pl}{\pl \varepsilon} f(u+\varepsilon u^{k+1})
\sqrt{d(u+\varepsilon u^{k+1})} \mid_{\varepsilon=0}\\
&=&\frac{\pl}{\pl \varepsilon} \Bigl(f(u)+\varepsilon u^{k+1} f'(u)\Bigr)
\Bigl(1+\frac{\varepsilon}{2}(k+1) u^k\Bigr)\sqrt{du}\mid_{\varepsilon=0} \\
&=&\frac{\pl}{\pl \varepsilon} \Biggl(f(u)\sqrt{du} +\varepsilon \Bigl(u^{k+1}
\frac{\pl}{\pl u} +\frac{k+1}{2} u^k\Bigr)f(u) \sqrt{du}\Biggr)_{\varepsilon=0} \\
&=&\sqrt{du}(u^{k+1}\frac{\pl}{\pl u} +\frac{k+1}{2} u^k) f(u) \\
&=&\frac{\sqrt{du}}{2}\Bigl( (u^{k+1} \frac{\pl}{\pl u} +\frac{\pl}{\pl u} u^{k+1})
f(u)\Bigr).
\end{eqnarray*}
and similarly for $u$ replaced by $v$. Setting $f=X(u,v)$ ends the proof of
Corollary 3.2.1.

\section{Virasoro equations satisfied by vertex operators and $\tau$-functions}

For distinct roots $\om,\om'\in\zeta_p$, the vertex operator
$X(t,\om z,\om'z)$ transforms the space of $\tau$-functions for the $p$-reduced
KP into itself and satisfies some infinite set of differential equations,
involving the Virasoro algebra; for the definition
of $W^{(2)}_{\ell}$, see footnote 3. This section is due to
Adler-Shiota-van Moerbeke \cite{ASV3}.

\proclaim Theorem 4.1.   For distinct $\om, \om' \in \zeta_p$ and for
$\ell \in\BZ$, $\ell \geq -p$ such that $p|\ell$, the
vertex operator $Y(z):=X(t,\om z, \om'z)$ satisfies the equations:
\be
\frac{\pl}{\pl z} z^{\ell +1}  Y(z)
= \left[\frac{1}{2}W^{(2)}_{\ell},Y(z)\right]
\label{vertexdiff}
\ee
\be
\left(-b^{\ell+1}\frac{\pl}{\pl b}-a^{\ell+1}\frac{\pl}{\pl
a}+\left[\frac{1}{2}W^{(2)}_{\ell},\,\cdot\,\right]\right)
\int^b_a dz Y(z)
=0.
\label{vertexint}
\ee
\be
\left(-b^{\ell+1}\frac{\pl}{\pl b}-a^{\ell+1}\frac{\pl}{\pl
a}+\left[\frac{1}{2}W^{(2)}_{\ell},\,\cdot\,\right]\right)
e^{-\lambda \int^b_a dz Y(z)}
=0,\label{35}
\ee

\medbreak
\noindent\underline{Proof}: At first we compute, using
Corollary 3.2.1,
$\om^p=\om'^{p} =1$ and $p | \ell$:
\begin{eqnarray*}
& &\frac{\pl}{\pl z}\left( z^{\ell+1} X(\om z,\om'
z) \right)\\
&=&\left(z^{\ell+1} \frac{\pl}{\pl z}+(\ell+1)z^{\ell}\right)X(\om z,\om'
z)\\
&=&\left(z^{\ell+1} \om
\frac{\pl}{\pl u}+z^{\ell+1} \om' \frac{\pl}{\pl
v}+(\ell+1)z^{\ell}\right)X(u,v)\mid_{{u=\om z}\atop{v=\om'z}}\\
&=&\left(u^{\ell+1}
\frac{\pl}{\pl u}+v^{\ell+1}\frac{\pl}{\pl
v}+(\ell+1)u^{\ell}\right)X(u,v)\mid_{{u=\om z}\atop{v=\om'z}}\\
&=&\left(v^{\ell+1}
\frac{\pl}{\pl v}+\frac{\pl}{\pl
u}u^{\ell+1}\right)X(u,v)\mid_{{u=\om
z}\atop{v=\om'z}}\\
&=&\frac{1}{2}\left[W^{(2)}_{\ell},X(u,v)\right]\mid_{{u=\om
z}\atop{v=\om'z}}\\
&=&\left[\frac{1}{2}W^{(2)}_{\ell},X(\om z,\om' z)\right],
\end{eqnarray*}
establishing (\ref{vertexdiff}). Upon using the above relation, one computes
\begin{eqnarray*}
& &\left(-b^{\ell+1}\frac{\pl}{\pl b}-a^{\ell+1}\frac{\pl}{\pl
a}+\frac{1}{2}\left[W_{\ell}^{(2)},\,\cdot\,\right]\right)
\int^b_a dz X(\om z,\om'z)\\
&=&-b^{\ell+1} X(\om b,\om' b) +a^{\ell+1} X(\om
a,\om' a) +\int^b_a dz \left[\frac{1}{2} W_{\ell}^{(2)} ,X(\om
z,\om'z)\right]\\
&=&-b^{\ell+1} X(\om b,\om' b) +a^{\ell+1}X(\om
a,\om' a)+z^{\ell+1} X(\om z,\om' z) |^b_a=0,
\end{eqnarray*}
establishing (\ref{vertexint}). Note
$$
V_{\ell}:=-b^{\ell+1}\frac{\pl}{\pl b}-a^{\ell+1}\frac{\pl}{\pl
a}+[\frac{1}{2} W_{\ell}^{(2)},\cdot]
$$
is a derivation. Therefore upon setting $U:=\int^b_a Y(z)dz$, expanding
the exponential and using $V_{\ell}U=0$, we have
$$
V_{\ell}(e^{-\lb
U})=V_{\ell}\left(\sum_0^{\iy}\frac{(-\lb)^n}{n!}U^n\right)=\sum_0^{\iy}
\frac{(-\lb)^n}{n!}\sum_{k=1}^n U^{n-k}(V_{\ell}U)U^{k-1}=0,
$$
establishing (39) and Theorem 4.1.

\bigbreak

\proclaim Corollary 4.1.1. The vertex operator
$$
Y(z):=e^{V(z)}X(t,\om z, \om'z)
$$
with distinct $\om, \om' \in \zeta_p$ satisfies the equations:
\be
\frac{\pl}{\pl z} z^{\ell +1} f(z) Y(z)
= \left[\tilde W_{\ell}^{(2)},Y(z)\right] ,
\label{vertexdiff'}
\ee
\be
\left(-b^{\ell+1} f(b) \frac{\pl}{\pl b}-a^{\ell+1} f(a)
\frac{\pl}{\pl a}+\left[\tilde W_{\ell}^{(2)},\,\cdot\,\right]\right)
\int^b_a dz Y(z)=0.
\label{vertexint'}
\ee
\be
\left(-b^{\ell+1}f(b) \frac{\pl}{\pl b}-a^{\ell+1} f(a) \frac{\pl}{\pl
a}+\left[\tilde W_{\ell}^{(2)},\,\cdot\,\right]\right)
e^{-\lambda \int^b_a dz ~Y(z)}=0,
\label{35'}
\ee
where
$$
\tilde W^{(2)}_{\ell}:=\frac{1}{2}\sum_{{k\in
p\BZ}\atop{k\geq 0}}a_kW^{(2)}_{\ell +k}+\frac{1}{\om-\om'}\sum_{{k\in
p\BZ}\atop{k\geq 0}}b_kW^{(1)}_{\ell +k+1},~~\mbox{ for
$\ell \in\BZ$, $\ell \geq -p$ such that $p|\ell$}
$$
$$
V'=:\frac{g(z)}{f(z)}=\frac{\sum_{i\geq 0}b_{pi}z^{pi}}{\sum_{i\geq
0}^{\iy}a_{pi}z^{pi}}.
$$

\medbreak
\noindent\underline{Proof}:
\begin{eqnarray*}
& &\frac{\pl}{\pl z}z^{\ell+1}f(z)e^{V(z)}X(\om z,\om'z)\\
&=&\left(\frac{\pl V}{\pl z}(z)f(z)\right)z^{\ell+1}X(\om
z,\om'z)e^{V(z)}+e^{V(z)}\frac{\pl}{\pl z}(f(z)z^{\ell+1}X(\om z,\om'z))\\
&=&\left(\sum_{k\in p\BZ}b_kz^{k+\ell+1}X(\om
z,\om'z)\right)e^{V(z)}+e^{V(z)}\frac{\pl}{\pl z}\left(\sum_{k\in
p\BZ}a_kz^{k+\ell+1}X(\om z,\om'z)\right)\\
&=&\left[\frac{1}{\om-\om'}\sum_{k\in p\BZ}b_k W^{(1)}_{k+\ell+1}\,\,\,
,\,\,\, X(\om z,\om'z)e^{V(z)}\right]\quad\mbox{(by (\ref{cor1}) and}\\ &
&\hspace{8cm}\om^{k+\ell+1}=\om\\ & &\hspace{8cm}\om'^{k+\ell+1}=\om')\\ &
&+\frac{1}{2}\left[\sum_{k\in p\BZ}a_kW_{k+\ell}^{(2)}\,\,\,
,\,\,\, X(\om
z,\om'z)e^{V(z)}\right]\quad\mbox{(by (\ref{cor2})),}
\end{eqnarray*}
concluding the proof of relation (\ref{vertexdiff'}) while the two relations
following (\ref{vertexint'}) and (\ref{35'}) are proved exactly in the way
the relations (\ref{vertexint}) and (\ref{35}) are derived from
(\ref{vertexdiff}).

The next theorem shows that if a function $f$ satisfies a Virasoro constraint,
then the Virasoro vector fields acting on the function
$$e^{-\lambda \int^b_a dz \,X(\om z,\om'z)}f,$$
move the end points $a$ and $b$ according to the simple vector field
\bea
\dot{a} =a^{kp+1} \mbox{   and   }\dot{b} =b^{kp+1}.\label{A}
\eea

\proclaim Theorem 4.2. For an arbitrary function $f(t)$ satisfying for some
$k\geq -1$
$$
W_{kp}^{(2)}f=c_{kp}f,
$$
we have for that same $k$
$$
\left(-b^{kp+1}\frac{\pl}{\pl b}-a^{kp+1}\frac{\pl}{\pl
a}+\frac{1}{2}W_{kp}^{(2)}-\frac{1}{2}c_{kp}\right)
e^{-\lambda \int^b_a dz\, X(\om z,\om'z)}f=0.
$$

\medbreak
\noindent\underline{Proof}: Setting, as before, $U=\int_a^bdz \, X(\om
z,\om'z)$, it follows from Theorem 4.1 that
\begin{eqnarray*}
0&=&\left(-b^{kp+1}\frac{\pl}{\pl b}-a^{kp+1}\frac{\pl}{\pl
a}+\frac{1}{2}\left[W_{kp}^{(2)},\,\cdot\,\right]\right)e^{-\lambda U} f(t) \\
&=&\left(-b^{kp+1}\frac{\pl}{\pl b}-a^{kp+1}\frac{\pl}{\pl
a}+\frac{1}{2}W_{kp}^{(2)}\right)e^{-\lambda U} f(t)-\frac{1}{2}
e^{-\lambda U}W_{kp}^{(2)}f(t) \\
&=&\left(-b^{kp+1}\frac{\pl}{\pl b}-a^{kp+1}\frac{\pl}{\pl
a}+\frac{1}{2}W_{kp}^{(2)}-\frac{1}{2} c_{kp}\right)e^{-\lambda U} f(t),
\end{eqnarray*}
using in the last equality the hypothesis, thus establishing Theorem 4.2.

\section{Fay  identities and symmetries}

The bilinear identity (\ref{bilinear}) generates many identities
for the
$\tau$-function. In the next proposition we describe some of them,
which will be useful in the next section:

\proclaim Proposition 5.1. (Fay identities). The $\tau$-function satisfies
the following bilinear identity
\be
\sum_{
\scriptstyle\rm cyclic\ permutations\atop
\scriptstyle\rm of\ indices\ \{1,2,3\}
}
(s_0-s_1)(s_2-s_3)\tau (t + [s_0] + [s_1])\tau(t + [s_2] + [s_3])
= 0.
\label{Fay}
\ee
and a more general identity
\begin{eqnarray}
& &\det\Bigl(\frac{\tau(t+[y_i^{-1}]-[z_j^{-1}])}
{\tau(t)}\frac{1}{y_i-z_j}\Bigr)_{1\leq
i,j\leq n} \nonumber\\
&=&(-1)^{\frac{n(n-1)}{2}}\frac{\Delta(y)\Delta(z)}{\prod_{k,\ell}
(y_k-z_{\ell})}\frac{\tau\Bigl(t+\sum_1^n[y_i^{-1}]-
\sum_1^n[z_j^{-1}]\Bigr)}{\tau(t)}.
\label{FayH}
\end{eqnarray}

\medbreak
\noindent\underline{Proof}: Upon shifting $t$ appropriately and using the
residue theorem, the bilinear identity (18) leads to the quadratic Fay identity.
It is also the simplest case of the Pl\"ucker relations for the corresponding
infinite-dimensional plane (see \cite{Sato,Shiota,vM}).

\bigbreak

The proof of the second relation (\ref{FayH}) proceeds by induction and is
due to \cite{ASV3}; at first, it is based on the bilinear relation
(\ref{bilinear}) for
$$
t\mapsto t+[y_1^{-1}]\quad\mbox{and}\quad t'\mapsto
t+\sum^n_{k=2}[y_k^{-1}]-\sum^n_{\ell=1}[z_{\ell}^{-1}].
$$
Using (\ref{log}), one computes
\bea
0&=&\oint\tau(t+[y_1^{-1}]-[z^{-1}])\,\,\tau\left(t+\sum_2^n[y_k^{-1}]
-\sum^n_{\ell=1}[z_{\ell}^{-1}]+[z^{-1}]\right)\nonumber\\
& &\hspace{6cm}e^{\sum_{i=1}^{\iy}\left(
\frac{y_1^{-i}}{i}-\sum^n_{k=2}\frac{y_k^{-i}}{i}+\sum^n_{\ell=1}
\frac{z_{\ell}^{-i}}{i}\right){z^i}}dz\nonumber\\
&=&\oint\tau(t+[y_1^{-1}]-[z^{-1}])\,\,\tau\left(t+\sum_{k=2}^n[y_k^{-1}]
-\sum^n_{\ell=1}[z_{\ell}^{-1}]+[z^{-1}]\right)\nonumber\\
& &\hspace{7cm}\frac{\prod^n_{\alpha =2}\left(1-\frac{z}{y_{\al}}\right)}
{\left(1-\frac{z}{y_1}\right)
\prod^n_{\beta =1}\left(1-\frac{z}{z_{\beta}}\right)}dz\nonumber\\
&=&-\tau(t)\,\,\tau\left(t+\sum^n_{k=1}[y_k^{-1}]-\sum^n_{\ell=1}
[z_{\ell}^{-1}]\right)\frac{y_1\prod^n_{\alpha =2}
\left(1-\frac{y_1}{y_{\alpha}}\right)}
{\prod^n_{\beta =1}\left(1-\frac{y_1}{z_{\beta}}\right)}\nonumber\\
& &-\sum^n_{\gamma =1}\tau(t+[y_1^{-1}]-[z_{\gamma}^{-1}])\,\,\tau
\left(t+\sum_{k=2}^n[y_k^{-1}]
-\sum_{{\ell=1}\atop{\ell\neq\gamma}}^n
[z_{\ell}^{-1}]\right)\frac{z_{\gamma}\prod^n_{\alpha =2}
\left(1-\frac{z_{\gamma}}{y_{\alpha}}\right)}
{\left(1-\frac{z_{\gamma}}{y_1}\right)
\prod_{{\beta =1}\atop{\beta\neq\gamma}}^n
\left(1-\frac{z_{\gamma}}{z_{\beta}}\right)}\nonumber\\
& &\label{41}
\eea

Develop now the determinant on the left hand side of (\ref{FayH}) according
to the first row and assume the validity of (\ref{FayH}) for $n\mapsto
n-1$; we find
\begin{eqnarray*}
& &\sum^n_{j=1}(-1)^{j-1}\frac{\tau(t+[y_1^{-1}]-[z_j^{-1}])}{\tau(t)}
\frac{1}{y_1-z_j}\det\left(\frac{\tau(t+[y_i^{-1}]-[z_k^{-1}])}{\tau(t)}
\frac{1}{y_i-z_k}
\right)_{\begin{array}{c}
2\leq i\leq n\\
1\leq k\leq n\\
k\neq j
\end{array}}\\
&=&\sum^n_{j=1}(-1)^{j-1}\frac{\tau(t+[y_1^{-1}]-[z_j^{-1}])}{\tau(t)}
\frac{1}{y_1-z_j}(-1)^{\frac{(n-1)(n-2)}{2}}\frac{\Dt(y_2,...,y_n)\Dt(z_1,...,\hat
z_k,...,z_n)}{\prod_{{k\neq 1}\atop{\ell\neq j}}(y_k-z_{\ell})}\\
&
&\hspace{7cm}\frac{\tau\left(t+\sum_2^n[y_i^{-1}]-\sum_{k\neq
j}[z^{-1}_k]\right)}{\tau(t)}\\
&=&(-1)^{\frac{n(n-1)}{2}}\frac{\Dt(y)\Dt(z)}{\prod_{k,\ell}(y_k-z_{\ell})}
\frac{\tau\left(t+\sum_1^n[y_i^{-1}]-\sum_1^n[z_j^{-1}]\right)}{\tau(t)}
\end{eqnarray*}
using the identity (\ref{41}); note for $n=1$, relation (\ref{FayH}) reduces
to (\ref{Fay}),  ending the proof of Proposition 5.1.

\bigbreak

If $\tau$ is a $\tau$-function, then
$X(y,z)\tau$ is also a
$\tau$-function, but also
$$
e^{aX(y,z)}\tau=\tau+aX(y,z)\tau,
$$
as follows from (\ref{vertexexp}).
But more is true, and it is an important feature of the vertex operator:
$X(y,z)$ is a {\em generating function of the infinitesimal symmetries of the
manifold of $\tau$-functions} \cite{Date}.

A {\em generating function for the symmetry vector fields on the manifold
of wave functions} $\Psi(x,t,z)$ is given by
$$
\BY_N\Psi=-N_-\Psi~, \mbox{  with  }
N(z,y):=e^{(z-y)M}\delta\left(y,L\right)\,.
$$
where $N$ is the pseudo-differential operator corresponding to $n$, defined
in (29), by means of the relation $n\Psi=N\Psi$; i.e., $z$ and $d/dz$ in $n$
get replaced, in reverse order by $L$ and $M$ defined in (\ref{M}). Note the
operator identity
\begin{equation}
N(z,y)_-=-\Psi(x,t,z)\circ D^{-1}\circ\Psi^*(x,t,y).
\label{N-}
\end{equation}
Upon using the shift
operator\footnote{$e^{-\eta}f(t)=f(t-[z])$}
~$\eta=\eta(z)=\sum_1^\infty(1/i)z^{-i}\pl/\pl t_i$, the relation between the
infinitesimal symmetries represented by $X$ on the space of $\tau$-functions
and the ones represented by $\BY_N$ on the space of wave functions is given by
(see \cite{ASV1} and, for generalizations, \cite{VDL}).

\proclaim Lemma 5.2. (ASV identity).

\begin{equation}
{\BY_N\Psi\over\Psi}=(e^{-\eta}-1){X\tau\over\tau},
\label{ASV}
\end{equation}
where $\Psi=\Psi(t,z)$, $\tau = \tau(t)$, $N=N(u,y)$ and $X=X(y,u)$.

\section{Determinant of kernels and vertex operators}

The main contention of this section is Theorem 6.1; at first, it claims that
the kernel
$$D^{-1}(\Psi^*(x,t,y)\Psi(x,t,z))$$
with $D^{-1}$ interpreted as in
section 2, is the vertex operator applied to the $\tau$-function
divided by $\tau$. Secondly, the determinant of a matrix involving this
kernel equals the vertex operator iterated several times as in
(\ref{vertexrel}). Theorems 6.1 and 6.2 are due to \cite{ASV3}.

\proclaim Theorem 6.1. (KP) \newline (i)
\be
D^{-1}(\Psi^*(x,t,y)\Psi(x,t,z))=\frac{1}{\tau(t)}
X(t,y,z)
\tau(t).\label{kernel}
\ee
(ii) Assuming $|y_1|,|z_1|<|y_2|,|z_2|<\cdots<|y_n|,|z_n|$,
we have the following vertex operators relation
\begin{eqnarray}
\det\left(D^{-1}\Psi^*(x,t,y_i)\Psi(x,t,z_j)\right)_{1 \leq i,j \leq
n}&=&  \det\left(\frac{1}{\tau(t)}~X(t,y_i,z_j)
\tau(t)\right)_{1 \leq i,j \leq n} \\
&=&\frac{1}{\tau}\,X(t,y_1,z_1)... X(t,y_n,z_n) \tau \nonumber\\
\label{det}
\end{eqnarray}

\medbreak
\noindent\underline{Proof}: The identity in (i) established in \cite{AvM1} can
be obtained at once by taking a limit of identity (\ref{ASV}), when $u
\rightarrow z$. Indeed, from (\ref{N-}) and the prior identity, we have for the
left hand side of (\ref{ASV}),
$$
\lim_{u \rightarrow z} \frac{\BY_N \Psi(x,t,z)}{\Psi(x,t,z)}
=-D^{-1} \Psi^{\ast}(x,t,y) \Psi(x,t,z)
$$
and for the right hand side~~\footnote{use $e^{-\eta} of (\ref{ASV}),
X(t,y,u)=\frac{z-u}{z-y}X(t,y,u)$}, one finds, upon using the explicit formula
(4) for the vertex operator $X(t,y,u)$ and (\ref{log})
\begin{eqnarray*}
\lim_{u\rightarrow z} (e^{-\eta }-1)\frac{X(t,y,u)\tau(t)}{\tau(t)}
&=&\lim_{u\rightarrow z}
\left( \frac{z-u}{z-y}\frac{X(t,y,u)e^{-\eta}\tau}{e^{-\eta}\tau}-
\frac{X(t,y,u)\tau(t)}{\tau(t)}   \right)\\
&=&-\frac{X(t,y,z)\tau(t)}{\tau(t)},
\end{eqnarray*}
thus leading to the desired identity.

Another proof is based on Fay's trisecant identity (\ref{Fay}); indeed
differentiating
(\ref{Fay}) with regard to $s_0$, setting $s_0=s_3=0$, dividing by
$s_1s_2$ and shifting $t\cv t-[s_2]$, yields~~\footnote{
        $\{ f,g \} = \Wronskian[g,f] = f'g-fg'$.
}
\begin{eqnarray*}
\lefteqn{
\{\tau(t),\tau(t+[s_1]-[s_2])\}
}\\
&&+(s_1^{-1}-s_2^{-1})\left(\tau(t+[s_1]-
[s_2])\tau(t)-\tau(t+[s_1])\tau(t-[s_2])\right)=0.
\end{eqnarray*}
Setting $y =s_1^{-1}$, $z =s_2^{-1}$ and
multiplying with $\tau(t)^{-2}\exp\sum t_i(z^i-y^i)$ lead to
$$
\frac{\tau(t+[y^{-1}])}{\tau(t)}
e^{-\sum_1^{\iy}t_jy^j} \frac{\tau(t-[z^{-1}])}
{\tau(t)}e^{\sum_1^{\iy}t_jz^j}=\frac{1}{z-y}
\frac{\pl}{\pl x}\left(
e^{\Sg t_i(z^i-y^i)}\frac{\tau(t+
[y^{-1}]-[z^{-1}])}{\tau(t)}\right),
$$
which implies (i).

To prove (\ref{det}), multiply
the higher Fay identity (\ref{FayH}) with appropriate exponentials, which
leads to  the following vertex operators relation:
\begin{eqnarray*}
\det\left(\frac{1}{\tau(t)}~X(t,y_i,z_j)
\tau(t)\right)_{1 \leq i,j \leq
n}
&=&(-1)^{\frac{n(n-1)}{2}}
\Delta(z)\Delta(y)
\frac{X(y_1,...,y_n;z_1,...,z_n)\tau}{\tau}\\
&=&\frac{1}{\tau}~X(t,y_1;z_1)...X(t,y_n;z_n) \tau,
\end{eqnarray*}
using in the last equality the identity (\ref{vertexrel}),
which is valid under the assumption $|y_1|,|z_1|<|y_2|,|z_2|<...$; upon using
(\ref{kernel}) in the left hand side of the above identity, one finds
(\ref{det}),ending the proof of Theorem 6.1.

\bigbreak

For future applications, we consider more general kernels of the form
$$
D^{-1}\sum_{\om\in\zeta_p}a_{\om}\Psi^*(x,t,\om
y)\sum_{\om'\in\zeta_p}b_{\om'}\Psi(x,t,\om'z),
$$
involving coefficients $a_{\om}$ and $b_{\om}\in\BC$, $\om\in\zeta_p$,
satisfying
$$
\sum_{\om\in\zeta_p}\frac{a_{\om}b_{\om}}{\om}=0.
$$
This kernel can  again be expressed in terms of a basic vertex operator
$$
Y(y,z):=\sum_{\om,\om'\in\zeta_p}a_{\om}b_{\om'}X(t,\om
y,\om'z)
$$
leading to a theorem, completely analogous to Theorem 6.1.

\proclaim Theorem 6.2. {(\bf $p$-reduced KP)}\newline (i)
\be
D^{-1}\sum_{\om\in\zeta_p}a_{\om}\Psi^*(x,t,\om y)\sum_{\om'\in\zeta_p}
b_{\om'}\Psi(x,t, \om'z)=\frac{1}{\tau}Y(y,z)\tau.
\label{46}
\ee
(ii) If $|y_1|,|z_1|\leq |y_2|,|z_2|\leq ...$, then
\newpage
\begin{eqnarray}
\det\left(D^{-1}(\sum_{\om\in\zeta_p}a_{\om}\Psi^*(t,\om
y_i)(\sum_{\om'\in\zeta_p}b_{\om'}\Psi(t,\om'z_j))\right)_{1\leq i,j\leq
n}&=&\det\left(\frac{1}{\tau}~Y(y_i,z_j)
\tau\right)_{1 \leq i,j \leq
n}.\nonumber\\
&=& \frac{1}{\tau}\left(\prod^n_{{i=1}\atop{{\rm ordered}}}
Y(y_i,z_i)\right)\tau \nonumber \\
\label{47}
\end{eqnarray}
If
\be
\sum_{\om\in\zeta_p}\frac{a_{\om}b_{\om}}{\om}=0,
\label{48}
\ee
the right hand side of (\ref{47}) has no singularities in the positive
quadrant $\{y_i\geq 0$ and $z_j\geq 0$ with $i,j=1,...,n\}$ and $\lim_{y\rg
z}Y(y,z)$ exists.

\medbreak
\noindent\underline{Proof of Theorem 6.2}: The proof of (i) follows at once
from Theorem 6.1. To prove (ii), we set
$$
k(y,z):=\frac{X(t,y,z)\tau(t)}{\tau(t)}=D^{-1}\Psi^*(t,y)\Psi(t,z);
$$
we have~~\footnote{$\vec\om_ky=(\om_{k1}y_1,...,\om_{kn}y_n)$}
\begin{eqnarray}
& &\det\left(D^{-1}(\sum_{\om\in\zeta_p}a_{\om}
\Psi^*(t,\om y_i)\Bigr)\Bigl(\sum_{\om'\in\zeta_p}b_{\om'}
\Psi(t,\om'z_j))
\right)_{1\leq i,j\leq n}\nonumber\\
&=&\det\Bigl(\sum_{\om,\om'\in\zeta_p}a_{\om}b_{\om'}k(\om
y_i,\om'z_j)\Bigr)_{1\leq i,j\leq n}\nonumber\\
&=&\sum_{\sg\in\pi_n}(-1)^{\vr(\sg)}\prod^n_{i=1}\Bigl(
\sum_{\om,\om'\in\zeta_p}a_{\om}b_{\om'}k(\om
y_i,\om'z_{\sg_i})\Bigr)\nonumber\\
&=&\renewcommand{\arraystretch}{0.5}
\begin{array}[t]{c}
\sum \\
{\scriptstyle \sg\in\pi_n}\\
{\scriptstyle \stackrel{\rg}{\om},\stackrel{\rg}{\om}'\in(\zeta_p)^n}
\end{array}
\renewcommand{\arraystretch}{1}(-1)^{\vr(\sg)}\prod_{1\leq i\leq n}
a_{(\stackrel{\rg}{\om})_i}b_{(\stackrel{\rg}{\om}')_i}k
((\stackrel{\rg}{\om})_iy_i,
(\stackrel{\rg}{\om}')_iz_{\sg_i})\nonumber\\
& &\quad\quad\mbox{upon interchanging
product\,}\prod^n_{i=1}\mbox{\,and sum\,}\sum_{\om,\om'\in\zeta_p}\nonumber\\
&=&\renewcommand{\arraystretch}{0.5}
\begin{array}[t]{c}
\sum \\
{\scriptstyle \sg\in\pi_n}\\
{\scriptstyle \stackrel{\rg}{\om},\stackrel{\rg}{\om}'\in(\zeta_p)^n}
\end{array}
\renewcommand{\arraystretch}{1}(-1)^{\vr(\sg)}\prod_{i=1}
a_{(\stackrel{\rg}{\om})_i}b_{(\sg\stackrel{\rg}{\om}')_i}
k((\stackrel{\rg}{\om})_iy_i,
(\sg\stackrel{\rg}{\om}')_iz_{\sg_i})\nonumber\\
& &\quad\quad\mbox{upon
replacing\,}\stackrel{\rg}{\om}'\in(\zeta_p)^n\mbox{\,by\,}\sg\stackrel{\rg}{\om}'
\in(\zeta_p)^n;\nonumber\\
&
&\quad\quad\mbox{note\,}(\sg\stackrel{\rg}{\om}')_i
=(\stackrel{\rg}{\om}')_{\sg_i}\nonumber\\
&=&\sum_{\stackrel{\rg}{\om},\stackrel{\rg}{\om}'\in(\zeta_p)^n}\det
\Bigl(a_{(\stackrel{\rg}{\om})_i}
b_{(\stackrel{\rg}{\om}')_j}k
((\stackrel{\rg}{\om})_iy_i,(\stackrel{\rg}{\om}')_jz_j)\Bigr)_{1\leq i,j\leq n}
\nonumber\\
&=&\sum_{\stackrel{\rg}{\om},\stackrel{\rg}{\om}'\in(\zeta_p)^n}\prod^n_{i=1}
a_{(\stackrel{\rg}{\om})_i}\prod^n_{j=1}
b_{(\stackrel{\rg}{\om}')_j}
\det(k((\stackrel{\rg}{\om})_iy_i,(\stackrel{\rg}{\om}')_jz_j)
\Bigr)_{1\leq i,j\leq n}\nonumber\\
&=&\frac{1}{\tau}\sum_{\stackrel{\rg}{\om},\stackrel{\rg}{\om}'\in(\zeta_p)^n}
\prod^n_{i=1}
a_{(\stackrel{\rg}{\om})_i}\prod^n_{j=1}
b_{(\stackrel{\rg}{\om}')_j}\Bigl(\prod^n_{{i=1}\atop{{\rm ordered}}}
X((\stackrel{\rg}{\om})_iy_i,(\stackrel{\rg}{\om}')_iz_i)\Bigr)\tau\nonumber\\
& &\quad\quad\mbox{\,for\,}|y_1|,|z_1|
\leq |y_2|,|z_2|\leq
...,\mbox{\,using\,}|({\stackrel{\rg}{\om}})_i|=1\nonumber\\
&=&\frac{1}{\tau}\Bigl(\prod^n_{{i=1}\atop{{\rm ordered}}}
(\sum_{\om,\om'\in\zeta_p}a_{\om}b_{\om'}X(\om
y_i,\om'z_i))\Bigr)\tau\nonumber\\
&=&\frac{1}{\tau}Y(y_1,z_1)Y(y_2,z_2)...Y(y_n,z_n)\tau
\nonumber
\end{eqnarray}
ending the proof of (\ref{47}).

We now prove that the expression above
\begin{eqnarray}
& &\tau\det\left(D^{-1}\sum_{\om\in\zeta_p}a_{\om}\Psi^*(t,\om
y_i)\sum_{\om'\in\zeta_p}b_{\om'}\Psi(t,\om'z_j)\right)_{1\leq i,j\leq
n}\nonumber\\
&=&\sum_{\vec\om_1,\vec\om_2\in(\zeta_p)^n}\prod^n_{i=1}a_{\om_{1,i}}
\prod^n_{j=1}b_{\om_{2,j}}\left(
\prod^n_{{i=1}\atop{\rm
ordered}}X((\vec\om_1)_iy_i,(\vec\om_2)_iz_i)\right)\tau\nonumber\\
&=&(-1)^{\frac{n(n-1)}{2}}\sum_{\vec\om_1,\vec\om_2\in
(\zeta_p)^n}\prod^n_{i=1}a_{\om_{1,i}}
\prod^n_{j=1}b_{\om_{2,j}}\frac{\Dt(\vec\om_2z)\Dt(\vec\om_1y)}{\prod_{k,\ell}
(\om_{2,k}z_k-\om_{1,\ell}y_{\ell})}\nonumber\\
& &\quad\quad\prod^n_{k=1}e^{\sum_{i=1}^{\iy}((\om_{2k}z_k)^i-
(\om_{1k}y_k)^i)t_i}\prod^n_{k=1}e^{\sum_{i=1}^{\iy}((\om_{1k}y_k)^{-i}-
(\om_{2k}z_k)^{-i})\frac{1}{i}\frac{\pl}{\pl t_i}}\tau,\label{49}
\end{eqnarray}
using Lemma 3.1 in the next to last equality, has no singularity in the positive
quadrant
$\{y_i\geq 0$ and
$z_j\geq 0$ with $i,j=1,...,n\}$. Indeed we compute the residue of this
function, near $y_{i_0}=z_{j_0}$. In the first sum
$\sum_{\vec\om_1,\vec\om_2\in(\zeta_p)^n}$, the only terms for which the
denominator blows up when $y_{i_0}=z_{j_0}$ are those for which
$\om_{1i_0}=\om_{2j_0}$; indeed in the denominator of (\ref{49}), the only
factor involving both $y_{i_0}$ and $z_{j_0}$ is
$\om_{2,j_0}z_{j_0}-\om_{1,i_0}y_{i_0}$, which vanishes at
$y_{i_0}=z_{j_0}$ if and only if $\om_{1i_0}=\om_{2j_0}$.

Setting $\om_{2j_0}=\om_{1i_0}$, the first fraction in (\ref{49})
behaves for $y_{i_0}$ near $z_{j_0}$, as follows:
\begin{eqnarray*}
&\,&\frac{\Dt(\stackrel{\rg}\om_1 y)\Dt(\stackrel{\rg}\om_2z)}
{\prod_{k,\ell}(\om_{2,k}z_k
-\om_{1,\ell}y_{\ell})}\\
&=&\frac{1}{\om_{2j_0}z_{j_0}-\om_{1i_0}y_{i_0}}\frac{\prod_{\ell
<i_0}(\om_{1\ell}y_{\ell}-\om_{1i_0}y_{i_0})\prod_{i_0<\ell}(\om_{1i_0}
y_{i_0}-\om_{1\ell}y_{\ell})}{\prod_{\ell\neq
i_0}(\om_{2j_0}z_{j_0}-\om_{1\ell}y_{\ell})}\\
&\,&\hspace{2cm}\times \frac{\prod_{i<j_0}(\om_{2i}z_i-\om_{2j_0}z_{j_0})\prod_{j_0<j}
(\om_{2j_0}z_{j_0}-\om_{2j}z_j)}{\prod_{k\neq
j_0}(\om_{2k}z_k-\om_{1i_0}y_{i_0})}\\
&\,&\hspace{2cm}\times\{\mbox{an
expression not involving\,}z_{j_0}\mbox{\,or\,}y_{i_0}\}\\
&&\\
&=&\frac{1}{\om_{1i_0}(z_{j_0}-y_{i_0})}\Biggl(\frac{\prod_{\ell
<i_0}(\om_{1\ell}y_{\ell}-\om_{1i_0}z_{j_0})\prod_{i_0<\ell}(\om_{1i_0}z_{j_0}
-\om_{1\ell}y_{\ell})}{\prod_{\ell\neq
i_0}(\om_{1i_0}z_{j_0}-\om_{1\ell}y_{\ell})} +O(y_{i_0}-
z_{j_0})\Biggr)\\
&\,&\Biggl(\frac{\prod_{i<j_0}(\om_{2i}z_i-\om_{2j_0}z_{j_0})\prod_{j_0<j}
(\om_{2j_0}z_{j_0}
-\om_{2j}z_j)}{\prod_{k\neq
j_0}(\om_{2k}z_k-\om_{2j_0}z_{j_0})} +O(y_{i_0}-
z_{j_0})\Biggr)\\
&\,&\times\{\mbox{expression not
involving\,}z_{j_0}\mbox{\,or\,}y_{i_0}\}\\
&&\\
&=&\frac{1}{\om_{1i_0}(z_{j_0}-y_{i_0})}(-1)^{i_0-1}(-1)^{n-j_0}\times
\left\{\begin{tabular}{c}
same expression not\\
involving $y_{i_0}$ or $z_{j_0}$ as above
\end{tabular}
\right\}+O(1).
\end{eqnarray*}
The second part in (\ref{49}) behaves as:
\begin{eqnarray*}
&=&\prod^n_{k=1}
e^{\sum_{i=1}^{\iy}((\om_{2k}z_k)^i-(\om_{1k}y_k)^i)t_i}\prod^n_{k=1}
e^{\sum_{i=1}^{\iy}((\om_{1k}y_k)^{-i}-(\om_{2k}z_k)^{-i})
\frac{1}{i}\frac{\pl}{\pl t_i}}\tau\\
&=&\prod_{k\neq
j_0}e^{\sum_{i=1}^{\iy}(\om_{2k}z_k)^it_i}\prod_{k\neq
i_0}e^{-\sum^{\iy}_{i=1}(\om_{1k}y_k)^it_i}\\
& & \hspace{3cm}\prod_{k\neq
i_0}e^{\sum^{\iy}_{i=1}(\om_{1k}y_k)^{-i}\frac{1}{i}\frac{\pl}{\pl
t_i}}\prod_{k\neq
j_0}e^{-\sum^{\iy}_{i=1}(\om_{2k}z_k)^{-i}\frac{1}{i}\frac{\pl}{\pl
t_i}}\tau
+O(z_{j_0}-y_{i_0})\\
&=&\prod^n_{k=1}
e^{\sum_{i=1}^{\iy}((\om_{2k}z_k)^i-(\om_{1k}y_k)^i)t_i}\prod^n_{k=1}
e^{\sum_{i=1}^{\iy}((\om_{1k}y_k)^{-i}-(\om_{2k}z_k)^{-i})
\frac{1}{i}\frac{\pl}{\pl
t_i}}\tau\Biggl|_{\stackrel{y_{i_0}=z_{j_0}}{w_{1,i_0 }=w_{2,j_0 } }}
 +O(z_{j_0}-y_{i_0}).
\end{eqnarray*}
Combining the two estimates, the sum of the terms in (\ref{49}),
corresponding to $\om_{1i_0}=\om_{2j_0}$, can be estimated as follows:
\begin{eqnarray*}
& &\renewcommand{\arraystretch}{0.5}
\begin{array}[t]{c}
\sum \\
{\scriptstyle \stackrel{\rg}{\om}_1,\stackrel{\rg}{\om}_2\in(\zeta_p)^n}\\
{\scriptstyle \mbox{with\,}\om_{1,i_0}=\om_{2,j_0}}
\end{array}
\renewcommand{\arraystretch}{1}\prod_{i=1}^na_{\om_{1,i}}
\prod_{j=1}^nb_{\om_{2,j}}
\Dt(\stackrel{\rg}{\om}_2z)\Dt(\stackrel{\rg}{\om}_1y)
X(\stackrel{\rg}{\om}_1y,\stackrel{\rg}{\om}_2z)\tau\\
&=&\renewcommand{\arraystretch}{0.5}
\begin{array}[t]{c}
\sum \\
{\scriptstyle
\hat{\stackrel{\rg}{\om}_1}:=(\om_{11},...,\hat\om_{1i_0},...,\om_{1,n})
\in(\zeta_p)^{n-1}}\\ {\scriptstyle
\hat{\stackrel{\rg}{\om}_2}:=(\om_{21},...,\hat\om_{2j_0},...,\om_{2,n})
\in(\zeta_p)^{n-1}}
\end{array}
\renewcommand{\arraystretch}{1} \sum_{\om_{1i_0}=\om_{2j_0}\in\zeta_p}\\
&
&\Biggl(\frac{1}{z_{j_0}-y_{i_0}}\frac{a_{\om_{1i_0}}b_{\om_{1i_0}}}
{\om_{1i_0}}\times\left\{
\begin{tabular}{c}
expression independent\\
of $y_{i_0}$ and $z_{j_0}$
\end{tabular}
\right\}\prod_{i\neq i_0}a_{\om_{1i}}
\prod_{j\neq j_0}b_{\om_{1j}}+O(1)\Biggr)\\
&=&\frac{1}{z_{j_0}-y_{i_0}}\sum_{\om\in\zeta_p}\frac{a_{\om}b_{\om}}{\om}
\sum_{\hat\om_1,\hat\om_2\in(\zeta_p)^{n-1}}\prod_{i\neq
i_0}a_{\om_{1i}}\prod_{j\neq j_0}b_{\om_{1j}}\left\{
\begin{tabular}{c}
expression independent\\
of $y_{i_0}$ and $z_{j_0}$
\end{tabular}
\right\}+O(1)\\
&=&O(1).
\end{eqnarray*}
using condition (\ref{48}).

To prove the last statement that $\lim_{y\rg z}Y(y,z)$ exists, note that, in
view of (4),
$$
X(y,z)=\frac{1}{z-y}+O(z-y).
$$
Therefore
\begin{eqnarray*}
Y(y,z)&=&\sum_{\om,\om'\in\zeta_p}a_{\om}b_{\om'}X(\om y,\om'z)\\
&=&\sum_{\om\in\zeta_p}a_{\om}b_{\om}X(\om y,\om z)+
\sum_{{\om,\om'\in\zeta_p}\atop{\om\neq\om'}}a_{\om}b_{\om'}X(\om y,\om'z)\\
&=&\left(\sum_{\om\in\zeta_p}\frac{a_{\om}b_{\om}}{\om}\right)\frac{1}{z-y}+O(1)\quad\quad\mbox{for
$z$ near $y$}\\ &=&O(1),
\end{eqnarray*}
ending the proof of Theorem 6.2.

\bigbreak

\section{Fredholm determinants and proof of Theorem 0.1}

The next point is to compute Fredholm determinants associated with the
kernels constructed in Theorems 6.1 and 6.2. First we need a formula
generalizing a classic soliton formula:

\proclaim Lemma 7.1.
$$\frac{1}{\tau}\prod^n_{
\scriptstyle1\atop
\scriptstyle\rm ordered
}
e^{a_iX(y_i,z_i)}\tau
=\frac{1}{\tau}\prod_1^n\left(1+a_iX(y_i,z_i)\right) \tau
=\det\left(\dt_{ij}+\frac{a_j}{\tau(t)}X(y_i,z_j)\tau(t)\right)_{1\leq
i,j\leq n}.
$$

\medbreak
\noindent\underline{Proof}:
\begin{eqnarray*}
&\,&\frac{1}{\tau} \prod_{{1}\atop{{\rm ordered}}}^n (1+a_i X(y_i,z_i)) \tau \\
&=&\frac{1}{\tau} (1+\sum_{1\leq i\leq n} a_i X(y_i,z_i) \tau + \sum_{1\leq
i<j\leq n} a_i a_j X(y_i,z_i) X(y_j,z_j)\tau + \cdots + \prod_1^n a_i \prod_1^n
X(y_i,z_i)\tau)\\
&=&1+\sum_{1\leq i\leq n} a_i\frac{X(y_i,z_i)\tau}{\tau} + \sum_{1\leq i< j
\leq n} \det \MAT{2}
a_i \frac{X(y_i,z_i)\tau}{\tau} & a_j \frac{X(y_i,z_j)\tau}{\tau}\\
a_i \frac{X(y_j,z_i)\tau}{\tau} & a_j \frac{X(y_j,z_j)\tau}{\tau}
\mat\\
& &\hspace{7cm}+ \cdots + \det (a_j \frac{X(y_i,z_i)\tau}{\tau})_{1\leq
i,j\leq n}\\ &=&\det (\delta_{ij} +
a_j\frac{X(y_i,z_i)\tau(t)}{\tau(t)})_{1\leq i,j\leq n}.
\end{eqnarray*}
This formula is well known in the context of soliton solutions to KdV and
KP, where it is applied to the $\tau$-function $\tau_0=1$; see for instance
\cite{Date,Ka,McK}. For KdV, we find the customary KdV $n$-soliton
solution, i.e.,
$$
q=2\frac{d^2}{dx^2}\log\tau(\t),
$$
with $\t=t+(x,0,0,\dots)$ as in Sect.~\ref{Three} and
$$
\tau(t)=\frac{1}{\tau_0}\left(\prod^n_{
\scriptstyle1\atop
\scriptstyle\rm ordered
}
e^{a_iX(y_i,-y_i)}
\right)\tau_0=\det\left(\dt_{ij}-\frac{a_j}{y_i+y_j}
e^{-\sum_{k:\rm odd}t_k(y^k_i+y_j^k)}\right)_{1\leq
i,j\leq n}.
$$

\bigbreak

The following is a continuous version of Lemma 7.1, where one replaces the
determinant appearing in Lemma 7.1 by a Fredholm determinant. This
statement is due to Adler-Shiota-van Moerbeke \cite{ASV3}.

\proclaim Theorem 7.2. Given the kernel
$$
k(y,z):=D^{-1}\Bigl((\sum_{\om\in\zeta_p}a_{\om}\Psi^*(t,\om y)(
\sum_{\om'\in\zeta_p}b_{\om'}\Psi(t,\om'z))\Bigr) = \frac{1}{\tau} Y (y,z)
\tau,
$$
and a subset
$E=\bigcup^r_{i=1}[a_{2i-1},a_{2i}]
\subset
\BR_+$, the Fredholm determinant of $k^E(y,z):=k(y,z)I_E(z)$ can be expressed
in terms of the original $\tau$-function:
\be
\det(I-\lb k^E)=\frac{1}{\tau}e^{-\lb\int_E dz\,Y(z,z)}\tau,
\label{50}
\ee
where
$$
Y(z,z):=\lim_{y\rg z}Y(y,z).
$$
If the $\tau$-function satisfies for some $k\geq -1$ the relation
$$
W_{kp}^{(2)}\tau=c_{kp}\tau,
$$
then $\tau\det(I-\lb k^E)$ satisfies for that same $k\geq -1$,
\be
\left(-\sum^{2r}_{i=1}a_i^{kp+1}\frac{\pl}{\pl
a_i}+\frac{1}{2}(W_{kp}^{(2)}-c_{kp})\right)\tau\det(I-\lb k^E)=0.\label{51}
\ee

\medbreak
\noindent\underline{Proof}:
We now apply the Fredholm determinant formula
\be
\det(I- \lb A)=1+\sum^{\infty}_{m=1} (-\lb)^m
\mathop{\int\!\ldots\!\int}\limits_{z_1\leq \cdots \leq z_m}
\det \left(A(z_i,z_j) \right)_{1\leq i,j\leq m}dz_1\ldots dz_m
\label{52}
\ee
to the kernel
$A=k^E (y,z):=k(y,z)I_E (z)$, for a given subset $E \subset \BR_+$. It
suffices to prove for $E = [a,b]$.  At first, the determinant appearing in
(\ref{52}) can be computed, according to (\ref{kernel}) and (\ref{det}) of
Theorem 6.1:
\be
F(z_1,\ldots,z_m):=\det(k(z_i, z_j))_{1\leq i,j \leq m} =
\frac{1}{\tau}\left(\prod^m_{{i=1}\atop{{\rm ordered}}} Y(z_i,z_i)\right)
\tau. \label{F}
\ee
We replace the integrals above by Riemann sums, with
$$
z_i=a+(i-1)\dt z, \dt z=\frac{b-a}{N-1},z_1=a, z_N=k, 1\leq i\leq N,
$$
so
\begin{eqnarray*}
\det(I-\lb k^E) &=&
\sum_{m:0}^{\iy} (-\lb)^m \int_{a<z_1<\cdots<z_m<b} \,\, F(z_1,\ldots,z_m)
\quad dz_1 \ldots dz_m\\
&=& \lim_{N\uparrow\iy} \sum^{N}_{m:0} (-\lb)^m \sum_{1\leq i_1<i_2<\cdots
<i_m \leq N}(\dt z)^m \,\, F(z_{i_1}, z_{i_2}, \ldots,z_{i_m}).
\end{eqnarray*}
We now compute the Riemann sums
\bea
& &\sum_{m:0}^N (-\lb)^m \sum_{1\leq i_1<i_2<\cdots <i_m\leq N} (\dt z)^m
\,\,F(z_{i_1},z_{i_2},\ldots,z_{i_m})\nonumber\\
&=&\sum^N_{m:0} \sum_{1\leq i_1<i_2<\cdots <i_m\leq N} (-\lb\,\dt z)^m
\frac{Y(z_{i_1}, z_{i_1})\cdots Y(z_{i_m}, z_{i_m}) \tau}{\tau},\mbox{using
(53)},\nonumber\\
&=&\frac{1}{\tau}\prod^N_{
\scriptstyle j:1\atop
\scriptstyle\rm ordered
}\bigl(1-\lb \,\dt z\,Y(z_j,z_j)\bigr)\tau,\mbox{\,using Lemma 7.1}\nonumber \\
&=&\frac{1}{\tau} \prod_{j:1}^N e^{-\lb\, \dt z
\,Y(z_i,z_i)} \tau, \mbox{ using (\ref{vertexexp})}\nonumber\\
&=&\frac{1}{\tau} e^{-\sum_{j:1}^N  \lb\, \dt z \,Y(z_i,z_i)} \tau,
\label{54}
\eea
using the fact that
$$
[Y(z_j,z_j), Y(z_m,z_m)]=0, \quad z_j< z_k;
$$
this is so because
$$
[X(t, \om z_j, \om'z_j), X(t,\om^{''}z_k, \om^{'''}z_k)]=0,
$$
since
$$
|\om z_j |<|\om^{'''}z_k| \mbox{ and } |\om'z_j|<|\om^{''} z_k|.
$$

Finally, taking the limit of the exponent in (\ref{54}), leads to
$$
\lim_{N\uparrow \iy} \sum_{j:1}^N \lb\, \dt z \,Y(z_i,z_i)=\lb \int_a^b dz \,
Y(z,z),
$$
establishing (\ref{50}). This fact together with Theorem 4.2 ends  the proof of
Theorem 7.2.

\proclaim Corollary 7.2.1. Given the kernel
$$
k(z,z')=D^{-1}\Psi^*(t,\om z)\Psi(t,\om'z')
$$
for two distinct roots $\om,\om'$ of $\om^p=1$, we have
$$
\det(I-\lb k^E)=\frac{1}{\tau}e^{-\lb\int_E dz \, X(t,\om z,\om'z)}\tau. $$

\medbreak
\noindent\underline{Proof}: Since $\om \neq \om'$, we have $a_{\om}=b_{\om'}=1$
and $a_{\om'}=b_{\om}=0$ and thus $\sum\frac{a_{\om}b_{\om}}{\om}=0$.
Therefore
$$
Y(y,z)=X(t,\om y,\om'z)\quad\mbox{and thus}\quad Y(z,z)=X(t,\om
z,\om'z),
$$
yielding the result.

\bigbreak

We now define a new kernel
\be
K(\lb,\lb'):=\frac{1}{p}\frac{k(z,z')}{z^{\frac{p-1}{2}}z'^{\frac{p-1}{2}}}
\label{55}
\ee
where $ \lb=z^p$ and $ \lb'=z'^p$.
For any $E\subset\BR_+$, let $K^E(\lb,\lb')=K(\lb,\lb')I_E(\lb')$.
Since
$$
K(\lb,\lb')\sqrt{d\lb} \sqrt{d\lb'}=k(z,z') \sqrt{dz}\sqrt{dz'},
$$
we have
$$
\det(I-K^E)=\det(I-k^{E^{1/p}}),
$$
where $E^{1/p}:=\{x\in\BR_+\mid x^p\in E\}$.
Thus from Theorem 7.2, we have:

\proclaim Corollary 7.2.2. For the kernel $K(\lb,\lb')$ and
a subset $E\subset\BR_+$, we have
\be
\det(I-\mu
K^{E})=\frac{1}{\tau}\left(e^{-\mu\int_{E^{1/p}}dzY(z,z)}\right)\tau.
\label{56}
\ee
Moreover, if $E$ is a finite union of intervals:
\be
E=\bigcup^r_{i=1}[A_{2i-1},A_{2i}],
\label{57}
\ee
and if $\tau(t)$ satisfies for $k\geq -1$,
$$
W_{kp}^{(2)}\tau=c_{kp}\tau,
$$
then for that same $k$,
\be
\left(- \sum^{2r}_{i=1}A_i^{k+1}\frac{\pl}{\pl A_i}
+\frac{1}{2p}(W_{kp}^{(2)}-c_{kp})\right)
\left(\tau\,\det (I-\mu K^{E})\right)=0.
\label{58}
\ee

\bigbreak

\noindent\underline{Proof of Theorem 0.1}: The first part of the statement
follows from Theorem 6.2, whereas part (ii) follows from Theorem 7.2.

\section{Virasoro and KP: a new system of PDE's}

To each $J_k^{(1)}$, $k\in\BZ$, we associate the weight $k$ and to a constant
the weight
$0$; thus
$$
{\rm weight}\left(c\prod^m_{i=1}J_{k_i}^{(1)}\right)=\sum_1^mk_i.
$$
Given the operator
$$
Q=\sum_{k_1,...,k_m\in\BZ}a_{k_1...k_m}J^{(1)}_{k_1}J^{(1)}_{k_2}
...J^{(1)}_{k_m}
$$
we define its ``principal symbol"
$$
\sg(Q)=\mbox{\,sum of terms in $Q$ of maximal weight}
$$
and the ``highest gap" $\gamma_Q$ is the difference between the maximal weight
and the next to maximal weight in $Q$.

In the next lemma, we consider $t$-operators of weight at most $k$ of the form
\be
Q_k=\sum_{i\leq k}a_{ki}J_i^{(1)}+\sum_{i+j\leq k}
b_{k,i+j}\colon J_i^{(1)}J_j^{(1)}\colon +d_k\
\mbox{ with } a_{kk}=1,
\label{60}
\ee
with a constant $d_k $, and differential operators ${\cal A}_k$ in parameters
$a$ of precise weight
$k$. We also need to give a precise meaning to
\be
p_i(\tilde\pl_{{\cal A}}):=p_i({\cal A}_1,\frac{1}{2}{\cal
A}_2,\frac{1}{3}{\cal A}_3,...);
\label{61}
\ee
the commutative variables $t_i$ get replaced by the
non-commutative operators $\frac{1}{i}{\cal A}_i$ with the following
ordering: if
$i_1\leq i_2\leq ...\leq i_{\ell}$, then
$$
t_{i_1}t_{i_2}...t_{i_{\ell}}\longmapsto\frac{1}{i_{\ell}...i_1}{\cal
A}_{i_{\ell}}...{\cal A}_{i_1}.
$$

\proclaim Lemma 8.1. Let a KdV $\tau$-function $\tau(t,a)$, depending on
$t\in\BC^{\iy}$ and an extra-set of parameters $a$, satisfy equations of the
type
\be
Q_k \tau={\cal A}_k\tau, \mbox{  with  }k=1,2,...;
\label{62}
\ee
with $Q_k$ as in (65), then the following
holds :
\medbreak
\noindent\underline{if $b_{k,k}=0$} for all $k=1,2,...$, then
$\tau(0,a)$ satisfies a hierarchy of bilinear PDE's, of the form
\bea
\tau\cdot{\cal A}_n{\cal A}_1\tau-{\cal A}_n\tau\cdot{\cal
A}_1\tau&-&\sum_{i+j=n+1}p_i(\tilde\pl_{{\cal A}})\tau\cdot
p_j(\tilde\pl_{{\cal A}})\tau\nonumber\\
& &+\mbox{(terms of weight $\leq n+1-\gamma$)}=0\label{63}
\eea
\hfill for $n=3,4,5,... .$
\medbreak\noindent If one shifts $t_1\mapsto t_1+x$ in the equations (67),
then $x$ figures in the lower-weight terms of (68).
\medbreak
\noindent\underline{if some $b_{k,k} \neq 0$} for some $k=1,2,...,$ and,
provided
 some determinant involving the $b_{k,k} $ is $\neq 0$, then $\tau(0,a)$
satisfies a similar hierarchy of PDE's:
\bea
\tau\cdot{\cal B}_{n1}\tau-{\cal B}_n\tau\cdot{\cal
B}_1\tau&-&\sum_{i+j=n+1}p_i(\tilde\pl_{{\cal B}})\tau\cdot
p_j(\tilde\pl_{{\cal B}})\tau\nonumber\\ & &+\mbox{(terms of weight $\leq
n+1-\gamma$)}=0\label{64}
\eea
\hfill for $n=3,4,5,...,$
\medbreak\noindent where $p_n(\tilde\pl_{{\cal B}})$, ${\cal B}_n$ and ${\cal
B}_{n-1,1}$ are appropriate linear combinations of
$$
{\cal A}_{j_{\ell}}...{\cal A}_{j_1},\quad\mbox{for\,\,}0\leq j_1\leq j_2\leq
...,\sum_{\alpha}j_{\alpha}=n,
$$
and
$$\gamma = \min_{k \in \Bbb{Z} } \gamma_{Q_k }.$$

\medbreak
\noindent\underline{Proof}: Consider the infinite set
$$\SR_k=\{k_1\geq k_2\geq ...\geq k_{\ell}\mbox{\,\,with\,\,} k_i\in\BZ
\mbox{\,\,and\,\,}\sum^{\ell}_{\al =1}k_{\al}=k\mbox{\,\,for arbitrary
$\ell$}\}
$$
and the finite set $\PR_k$ of partitions, defined by
\begin{eqnarray*}
\SR_k\supset\PR_k&=&\{k_1\geq k_2\geq ...\geq 0,\mbox{\,\,with\,\,} k_i\in\BZ
\mbox{\,\,and\,\,}\sum^{\ell}_1k_{\al}=k\mbox{\,\,for arbitrary
$\ell$}\}\\
&=&\{\mbox{Young diagrams $Y$ of weight $k$}\}.
\end{eqnarray*}
In terms of the ``principal symbol" $\sg(Q_k)$ of $Q_k$:
$$
\sg(Q_k)=J_k^{(1)}+b_{kk}\sum_{i+j=k}\colon J_i^{(1)}J_j^{(1)}\colon,\quad
k\geq 1,
$$
we have for given $Y=(k_1\geq k_2\geq ...\geq 0)\in\PR_k$
\bea
Q_{k_{\ell}}...Q_{k_1}&=&\sg(Q_{k_{\ell}})...\sg(Q_{k_1})+\mbox{\,lower
weight terms}\label{65}\\
&=&J_{k_{\ell}}^{(1)}...J_{k_1}^{(1)}+\sum_{(i_1\geq ...\geq
i_m)\in\SR_k}c^{i_1...i_m}_{k_1...k_{\ell}}J_{i_m}^{(1)}...J_{i_1}^{(1)}+
\mbox{\,lower weight terms,}\nonumber
\eea
with the $c^{i_1...i_m}_{k_1...k_{\ell}}$ being functions of the $b_{kk}$
only. For a given Young diagram $Y=(k_1\geq k_2\geq ...)$, we define
$$
J_Y^{(1)}:=J_{k_{\ell}}^{(1)}...J^{(1)}_{k_1};
$$
with this notation
\be
Q_{k_{\ell}}...Q_{k_1}\Bigl|_{t=0}=J_Y^{(1)}+\sum_{Y'\in\PR_k}c_Y^{Y'}J_{Y'}^{(1)}+
\mbox{\,lower weight terms + c.}
\ee
Thus, for given $k>0$, we get a system of $(\#\PR_k)$-equations in the
unknowns $J_Y^{(1)}$, $Y\in\PR_k$, which is invertible, provided
$$
\det\left(I+(c_Y^{Y'})_{Y',Y\in\PR_k}\right)\neq 0.
$$
Hence, for each $Y=(i_1\geq i_2\geq ...)\in\PR_k$ the equations (71) yield
\begin{eqnarray*}
J^{(1)}_{i_1}...J^{(1)}_{i_m}&=&\sum_{(j_1...j_{\ell})\in\PR_k}
d_{i_1...i_m}^{j_1...j_{\ell}}Q_{j_{\ell}}...
Q_{j_1}\Bigl|_{t=0}\\
& &+\left(\begin{array}{l}
\mbox{lower weight terms in $J_k^{(1)}$}\\
\mbox{with $k>0$}\end{array}\right)+\mbox{\,constant.}
\end{eqnarray*}
We now proceed by induction: for $k=1$ i.e., $\PR_1=\{(1)\}$, we have
$$Q_1=J_1^{(1)}+ \mbox{ (lower weight terms) }$$
 and thus
$$
J_1^{(1)}=\frac{\pl}{\pl t_1}=Q_1\Bigl|_{t=0}+\mbox{\,constant.}
$$
Therefore for each $Y=(i_1\geq i_2\geq ...\geq i_m)\in\PR_k$, $k\geq 1$,
$$
J^{(1)}_{i_1}...J_{i_m}^{(1)}=\sum_{{j_1\geq ...\geq
j_{\ell}\geq 1}\atop{1\leq j_1+...+j_{\ell}\leq k}}d_{i_1...i_m}
^{j_1...j_{\ell}}Q_{j_{\ell}}...
Q_{j_1}\Bigl|_{t=0}+\mbox{\,constant.}
$$
Therefore for a $\tau$-function satisfying (67), we have
\be
J^{(1)}_{i_1}...J_{i_m}^{(1)}\tau\Bigl|_{t=0}=\sum_{{j_1\geq ...\geq
j_{\ell}\geq 1}\atop{1\leq j_1+...+j_{\ell}\leq k}}d_{i_1...i_m}
^{j_1...j_{\ell}}{\cal A}_{j_1}...
{\cal A}_{j_{\ell}}\tau\Bigl|_{t=0}+\mbox{\,constant,}\label{66}
\ee
since, for instance,
$$Q_{j_2} Q_{j_1} \tau = Q_{j_2} {\cal A}_{j_1} \tau = {\cal A}_{j_1 }
Q_{j_2 } \tau = {\cal A}_{j_1} {\cal A}_{j_2} \tau.$$
Note $\tau$ also satisfies the $KP$ hierarchy
\be
\tau\cdot\frac{\pl^2\tau}{\pl t_k\pl
t_1}-\frac{\pl\tau}{\pl
t_k}\frac{\pl\tau}{\pl t_1}-\sum_{i+j=n+1}p_i(\tilde\pl)\tau\cdot
p_j(-\tilde\pl)\tau=0,\label{67}
\ee
\hfill $n=3,4,5,...$,
\medbreak\noindent whose terms all have the same weight $n+1$.

For instance, the expression $p_n(\tilde\pl)$ in the KP-equation, namely
$$
p_n\left(\frac{\pl}{\pl t_1},\frac{1}{2}\frac{\pl}{\pl
t_2},...\right)=\sum_{\renewcommand{\arraystretch}{0,5}
\begin{array}[t]{c}
k_i \geq 0\\
\sum_{i\geq 1} i k_i = n\\
\end{array}
\renewcommand{\arraystretch}{1}
}
\frac{(J_{\al}^{(1)})^{k_{\al}}...(J_{1}^{(1)})^{k_1}}{k_{\al}!...k_1!
\al^{k_{\al}}\ldots 1^{k_1 }},
$$ leads to a contribution
\begin{eqnarray*}
p_n(\tilde\pl_{{\cal B}})&=&\sum_{{k_i\geq 0}\atop{\sum_1^{\al}ik_i=n}}
\frac{\prod_{i}^{\alpha} r^{-k_r }}{k_{\al}!...k_1!}\sum_{{j_i\geq ...\geq
j_{\ell}\geq 1}\atop{\sum_i
j_i=n}}d^{j_1...j_{\ell}}_{\underbrace{\al,...,\al}_{k_{\al}},
\underbrace{\al-1,...,\al-1,...}_{k_{\al-1}}}{\cal A}_{j_1}...{\cal
A}_{j_{\ell}}\\
& &\hspace{6cm}+\,\mbox{\,lower weight terms}\\
&=&\sum_{{j_1\geq ...\geq j_{\ell}\geq 1}\atop{\sum_ij_i=n}}\left(
\sum_{{k_i\geq 0}\atop{\sum_1^{\al}ik_i=n}}
\frac{d^{j_1...j_{\ell}}_{\al,...,\al,\al-1,...,\al-1,...}}{\prod_{i}^{\alpha}
r^{k_r } k_{\al}!...k_1!}\right) {\cal A}_{j_1}...{\cal A}_{j_{\ell}}\\
& &\hspace{6cm}+\,\mbox{\,lower weight terms.}
\end{eqnarray*}
When $b_{k,k}=0$, for all $k=1,2,...$, then the proof is much simpler;
indeed, then formula (70) becomes
\be
Q_{k_{\ell}}...Q_{k_1}=J^{(1)}_{k_{\ell}}...J^{(1)}_{k_1}+
\mbox{\,lower weight terms in $J_k^{(1)}$},\label{68}
\ee
which leads at once to
$$
J_{k_{\ell}}^{(1)}...J_{k_1}^{(1)}\tau\Bigl|_{t=0}={\cal A}_{k_{\ell}}...{\cal
A}_{k_1}\tau\Bigl|_{t=0}+(\mbox{\,lower weight terms in ${\cal A}_k$}).
$$
Upon shifting $t_1 \mapsto t_1+x$, the variable $x$ will
only figure in the lower-weight terms of (65), using $b_{kk}=0$, and thus
$x$ appears in the lower-weight terms of (68) only. This ends the proof of Lemma
8.1.

\bigbreak

In the proof of Lemma 8.1, one needs to compute expressions like (\ref{65}): for
$(k_1\geq ...\geq k_{\ell})\in\PR_k$
$$
Q_{k_{\ell}}...Q_{k_1}\Bigl|_{t=0}=J_{k_{\ell}}^{(1)}...J^{(1)}_{k_1}+
\sum_{(i_1\geq ...\geq
i_m)\in\PR_k}c^{i_1...i_m}_{k_1...k_{\ell}}J_{i_m}^{(1)}...J^{(1)}_{i_1}+
\mbox{\,l.w.t.};
$$
i.e., typically, by virtue of the form (65) of $Q_k$, one is dealing with
products of normal ordered sums
$$
\left(\sum_{i+j=k_{\ell}}\colon J_i^{(1)}J_j^{(1)}\colon\right)...
\left(\sum_{i+j=k_1}\colon J_i^{(1)}J_j^{(1)}\colon\right)\,,
$$
which are then evaluated at $t=0$. They can actually be computed in terms of
graphs, as explained in Lemma 8.2.

Consider the following generating function
\begin{eqnarray*}
T(z)&:=&\colon J^{(1)}(z)J^{(1)}(z)\colon,
\mbox{\,\,with\,\,}J^{(1)}(z)=\sum_{i\in\BZ}J^{(1)}_iz^{-i-1}\mbox{\,\,and\,\,}
J_0^{(1)}=\mu_0\\
&=&\sum_{n\in\BZ}\left(\sum_{i+j=n}\colon
J_i^{(1)}J_j^{(1)}\colon\right)z^{-n-2}\\ &=&\sum_{n\in\BZ}J_n^{(2)}z^{-n-2}.
\end{eqnarray*}
Upon shifting $t_n\mapsto t_n+\mu_n$ /n
$$
J^{(1)}(z)=\sum_{i\in\BZ}J_i^{(1)}z^{-i-1}+\sum_{j\geq 1}\mu_jz^{j-1}
$$
and thus
\begin{eqnarray*}
J^{(1)}(z)\Bigl|_{t=0}&=&\sum_{{i\in\BZ}\atop{i\geq 1}}
J^{(1)}_iz^{-i-1}+\frac{\mu_0}{z}+\sum_{j\geq 1}\mu_jz^{j-1}=:J_0(z)\\
&=:&J_0(z).
\end{eqnarray*}
Also
$$
T(z)\Bigl|_{t=0}=J_0(z)^2.
$$

Consider the set of graphs
$$
\Gamma^{(n)}=\{G_{\al}^{(n)}\}
$$
where $G_{\al}^{(n)}$ is a graph with $n$ numbered vertices subjected to the
rule: out of every vertex, you have at most 2 links leading to another vertex.
Each graph $G_{\al}^{(n)}=\bigcup_i G_{\al i}^{(n)}$ consists of a number of
connected components as in the figure below
\bigbreak

$$
\stackrel{4}{\cdot}  \rule{10mm}{0mm}
\stackrel{2}{\cdot}  \rule{10mm}{.3mm}
\stackrel{3}{\cdot}  \rule{10mm}{.3mm}
\stackrel{8}{\cdot}  \rule{10mm}{.3mm}
\stackrel{7}{\cdot}  \rule{10mm}{.3mm}
\stackrel{6}{\cdot} \ ~~~~~
\stackrel{1}{\cdot}  \stackrel{\rule{10mm}{.2mm}}{\rule{10mm}{.2mm}}
\stackrel{5}{\cdot}
$$

\bigbreak

\noindent We will be using the following series:
$$
\sum_{i\geq 0}iz^{-i-1}w^{i-1}=\frac{1}{(z-w)^2}\mbox{\,\,and\,\,}
\sum_{i\geq 0}\frac{i(i^2-1)}{6}z^{-i-2}w^{i-2}=\frac{1}{(z-w)^4}.
$$

\proclaim Lemma 8.2.
$$
T(z_1)...T(z_n)=\sum_{G_{\al}^{(n)}\in\Gamma^{(n)}}\colon\prod_if_{G^{(n)}_{\al
i}}\colon
$$
where each $G^{(n)}_{\al i}$ contributes a factor in each term:
$$
\begin{array}{lll}
\mbox{if}&G^{(n)}_{\al i}=~~~\{k\}:&f_{G_{\al i}}=T(z_k)\\
&G^{(n)}_{\al i}=~~~\stackrel{\ell_1}{\cdot}
\rule{10mm}{.3mm} \stackrel{\ell_2}{\cdot} \rule{10mm}{.3mm}
\stackrel{\ell_3}{\cdot}\,
\cdots\,\cdot \rule{10mm}{.3mm} \stackrel{\ell_{\al}}{\cdot}:&f_{G_{\al
i}}=J^{(1)}(z_{\ell_1})
J^{(1)}(z_{\ell_{\alpha}})\prod^{\al-1}_{i=1}\frac{1}{(z_{\ell_i}-z_{\ell_{i+1}})^2}\\
&G^{(n)}_{\al i}=~~~\stackrel{m_1}{\cdot}
\stackrel{\rule{10mm}{.1mm}}{\rule{10mm}{.1mm}}
\stackrel{m_2}{\cdot}:&f_{G_{\al i}}=\frac{1}{2(z_{m_1}-z_{m_2})^4}
\end{array}
$$

\medbreak
\noindent\underline{Proof}: The proof proceeds by induction on $m$, the
relation being obviously true for $m=1$.

\proclaim Corollary 8.2.1. We have
$$
T(z_1)...T(z_m)\Bigl|_{t=0}=\sum_{G_{\al}^{(n)}\in\Gamma^{(n)}}\prod_if_{G_{\al
i}}\Bigl|_{t=0}
$$
where
$$
\begin{array}{lll}
\mbox{if}&G^{(n)}_{\al i}=~~~\{k\}:
&f_{G^{(n)}_{\al i}}\Bigl|_{t=0}=J_0^{(1)}(z_k)^2\\
&G^{(n)}_{\al i}=~~~\stackrel{\ell_1}{\cdot}
\rule{10mm}{.3mm} \stackrel{\ell_2}{\cdot} \rule{10mm}{.3mm}
\stackrel{\ell_3}{\cdot}\,
\cdots\,\cdot \rule{10mm}{.3mm} \stackrel{\ell_{\al}}{\cdot}:
&f_{G^{(n)}_{\al i}}\Bigl|_{t=0}=J_0^{(1)}(z_{\ell_1})
J_0^{(1)}(z_{\ell_{\al}})\prod^{\al-1}_{i=1}
\frac{1}{(z_{\ell_i}-z_{\ell_{i+1}})^2}\\
&G^{(n)}_{\al i}=~~~\stackrel{m_1}{\cdot}
\stackrel{\rule{10mm}{.1mm}}{\rule{10mm}{.1mm}}
\stackrel{m_2}{\cdot}:
&f_{G^{(n)}_{\al i}}\Bigl|_{t=0}=\frac{1}{2(z_{m_1}-z_{m_2})^4}
\end{array}
$$

\section{Proof of Theorem 0.2}
Remember from section 0 that
$$A = A_z = \frac{1}{2} z^{-m+1} (\frac{\partial}{\partial z} + V'(z) ) +
\sum_{i \geq 1} c_{-2i } z^{-2i } $$
with
$$V(z) = \frac{\alpha}{2} z + \frac{\beta}{6} z^3 \not\equiv 0, m = \deg V' = 0
\mbox{ or } 2.$$

\proclaim Lemma 9.1. Consider a function $\Psi (x,z)$ satisfying conditions (12)
and (13) of Theorem 0.2; i.e., $\Psi (x,z)$ is a solution of the linear partial
differential equation
\be
A_z \Psi (x,z) = \hat A_x \Psi (x,z)
\ee
with holomorphic $\Psi (0,z) \mbox{ in } z^{Ñ1} $, subjected to
\be
(a A^2 + bA + c) \Psi (0,z) = z^2 \Psi (0,z) ,~\mbox{ with } \Psi (0,z) = 1 +
O(z^{-1}). \ee
 Then $\Psi (x,z)$ specifies an infinite-dimensional plane
\footnote{$Gr^{(0)}$ is the big cell of the Grassmannian $Gr$.}
\be
W := \mbox{ span }_{\Bbb{C}}~ \{ \left(\frac{\partial}{\partial x}\right)^j \Psi
(x,z)|_{x=0} ,~~ j = 0,1,2,\ldots \} \in Gr^{0} (x),
\ee
invariant under $z^2 $ and $A$,
$$z^2 W \subset W \mbox{ and } AW \subset W,$$
such that
$$W^x := e^{-xz} W = \mbox{ spanÊ}~ \{ e^{-xz} \left(\frac{\partial}{\partial
x}\right)^j
\Psi (x,z), j = 0,1,2,\ldots \mbox{ for fixed } x \}.$$
with $\Psi (x,z) = e^{xz} (1 + O (z^{-1}) )$.  Also, there exists a potential
$q(x)$ such that
$$\left(\frac{d^2 }{dx^2 } + q(x)\right) \Psi (x,z) = z^2 \Psi (x,z).$$\par

\medbreak

\noindent \underline{Remark} : Another way of stating Lemma 9.1 goes as follows
: the tau-function $\tau(t)$ associated with
$$W^{x,t} = e^{-xz - \sum_1^{\iy} t_i z^i} W, \mbox{ with } W \mbox{ as in
} (77),$$
and the corresponding wave function
$$\Psi (x,t,z) = e^{xz + \sum_1^{\iy} t_i z^i } \frac{\tau (\bar t - [z^{-1}
])}{\tau (\bar t)}$$
are such that $\Psi (x,0,z) = \Psi (x,z).$

\bigbreak

\noindent \underline{Proof} : First observe that $(m = 0 \mbox{ or } 2)$:
\medbreak
\be
A_z = \frac{1}{2} z^{-m+1} \frac{\partial}{\partial z} + c_m z
+ O(1), \mbox{ with } c_0 = \frac{\alpha}{4} \mbox{ and } c_2
=\frac{\beta}{4} ,
\ee

\noindent with
\be
[A_z ,z^2 ] = z^{2-m} .
\ee

\bigbreak

Consider the function $\Psi (x,z)$, solution of (75) and (76) ; then the plane
$W \in Gr$, generated by $\Psi (0,z)$,
\be
W := \mbox{ span } \{ A^k_z \Psi (0,z), k = 0,1,2,\ldots\},
\ee
belongs to the big cell $Gr^{(0)}$, because
$$
A^k_z \Psi (0,z) = (c_m z)^k (1 + O(z^{-1})).
$$

\medbreak

\noindent The plan $W$ has the following property
$$A_z W \subset W \mbox{ and } z^2 W \subset W.$$

\medbreak

\noindent The first inclusion follows at one from the construction (80) of the
plane $W$, while the second requires some argument.  It suffices to prove
\be
z^2 A^k \Psi (0,z) = \sum_{i=0}^{k+2} c_i A^i \Psi (0,z),
\ee
which is done by induction.  The case $k = 0$ is merely condition (76) and we
assume (81) up to including $k$; we compute
$$
z^2 A^{k+1} \Psi (0,z)
\quad\quad\quad\quad\quad\quad\quad\quad\quad\quad\quad\quad\quad
\quad\quad\quad\quad\quad\quad\quad\quad\quad\quad\quad\quad\quad\quad
$$
$$
\begin{array}{ll}
    &= A^{k+1} z^2 \Psi (0,z) + [z^2 ,A^{k+1} ]\Psi (0,z)\\
    & \\
                  &= A^{k+1} z^2 \Psi (0,z) + \sum_{i=0}^k A^i [z^2 ,A] A^{k-i} \Psi(0,z)\\
    & \\
                  &= (aA^{k+3} + bA^{k+2} + cA^{k+1}) \Psi (0,z)
     - \sum_{i=0}^k A^i z^{2-m} A^{k-i} \Psi (0,z)\\
    & \quad \quad \quad \quad\quad\quad\quad\quad\quad
       \quad\quad\quad\quad\quad\quad\quad\quad\quad
       \mbox{ using (76) and (79).}\\
    &= \sum_{i=0}^{k+3} c_i A^i \Psi (0,z), \mbox{ for some } c_i .
\end{array}
$$
The process of letting $W$ evolve according to the $KP$-vector fields defines
a $\tau$-function and a wave function  $\Psi (x,t,z) = e^{xz + \sum t_i z^i }
(1 + O (z^{-1} ))$,
in terms of which the same plane $W$ can be expressed as
\be
W = \mbox{ span } \{ \left(\frac{\partial}{\partial x}\right)^j \Psi
(x,t,z),~~ j = 0,1,2,
\ldots \mbox{ for fixed } (x,t) \}. \quad
\ee
Therefore, whenever $AW \subset W$ for a differential operator $A = A_z $, we
have
\be
A \Psi (x,t,z) = \sum_{j \geq 0} c_j \left(\frac{\partial}{\partial x}\right)^j
\Psi (x,t,z).
\ee
Also, remember, to any operator $A \in \BC [z,z^{-1}
,\frac{\partial}{\partial z}Ê]$, we associate a pseudo-differential operator
$P_A $
$$
A = \sum_{\stackrel{-\infty < i < \infty}{0 \leq j < \infty}} c_{ij} z^i
\left(\frac{\partial}{\partial z}\right)^j \mapsto P_A = \sum_{\stackrel{-\infty
< i <
\infty}{0 \leq j < \infty}} c_{ij} M^j L^i,$$
such that
\be
A \Psi (x,t,z) = P_A \Psi (x,t,z).
\ee
Upon comparing (83) and (84), we have that $P_A $ is a differential operator for
that $\Psi$, i.e., $P_A = (P_A )_+ $.

Therefore, in view of the explicit form (9) of $A$ and the representation (23)
of $M$, we have
\be
\begin{array}{ll}
P_A &= \frac{1}{2} (M + V'(L)) L^{-m+1} + \sum_{i \geq 1} c_{-2i} L^{-2i}\\
    &  \\
    &= (\frac{1}{2} M L^{-m+1} + c_m L)_+ \\
    &   \\
    &= \left(S (\frac{1}{2} x D^{-m+1} + c_m D) S^{-1}\right)_+ + \frac{1}{2}
\sum_{k \geq 1} k t_k (L^{k-m})_+
\end{array}
\ee
and
\be
P_{z^2 } = L^2 = (L^2 )_+ = D^2 + q(x,t).
\ee
Setting $t = 0$ in (85) leads to
\begin{eqnarray}
P_A |_{t=0} &=& \left(S (\frac{1}{2} \, x D^{-m+1} + c_m D) S^{-1}\right)_+
\nonumber\\
   & & \nonumber \\
&= &((1 + a_{-1} D^{-1} + \ldots) (\frac{1}{2}Ê\, x D^{-m+1} + c_m D) (1
- a_{-1} D^{-1} + \ldots))_+ \nonumber  \\
  & & \nonumber \\
&= &\frac{\beta}{4} \delta_{m,2} D + \frac{1}{2} (x + \frac{\alpha}{2} )
\delta_{m,0} D = \hat A_x .
\end{eqnarray}
Therefore, combining (84) and (85), $\Psi (x,0,z)$ is a solution of the first
order partial differential equation
\be
A_z \Psi (x,0,z) = \hat A_x \Psi (x,0,z)
\ee
with
$$
\Psi (0,0,z) = \Psi (0,z),
$$
where $\Psi (0,z)$ satisfies (76).  But the function $\Psi (x,z)$,
alluded to in the statement of Lemma 9.1, satisfies, by assumption, the same
equation (75), with the same initial condition (76).  The solution of such a
linear partial differential equation is unique, because the relevant Jacobian
does not vanish, namely the coefficient of $D$ in $\hat A_x $.  Therefore we
conclude
$\Psi (x,0,z) = \Psi (x,z)$ and the plane (77) generated by $\Psi (x,z)$ has the
properties stated in Lemma 9.1.

\medbreak

\proclaim Lemma 9.2. A function $\Psi (x,z)$ satisfying conditions (12) and
(13) of Theorem 0.2, defines a unique $KdV \tau$-function $\tau (\bar t) =
\tau (x + t_1, t_2 ,\ldots)$ satisfying the Virasoro constraints
\be
\frac{1}{4} (W^{(2)}_k +Ê\alpha W^{(1)}_{k+1} + \beta W^{(1)}_{k+3}) = c_k
\tau \mbox{ for even } k \geq -m,
\ee
for some constants $c_k $.

\noindent\underline{Proof}: From the proof of Lemma 9.1, we have that
$$
P_{z^2} = L^2 ~\mbox{ and }~ P_A = \frac{1}{2} (ML^{-m+1} + \frac{1}{2} \alpha
L^{-m+1} + \frac{1}{2} \beta L^{-m+3})
$$
are differential operators and thus $(P_A L^{k+m} )_- = 0$ for all even $k \geq
-m$.  Then from the ASV-formula (see (48) in lemma 5.2):
$$\frac{-(M^n L^{n+\ell})_-  \Psi}{\Psi} = (e^{-\eta} - 1) \frac{\frac{1}{n+1}
W^{n+1}_{\ell} \tau}{\tau} $$
it follows that for all even $k \geq -m$
\begin{eqnarray*}
0 &=&  - \frac{(P_A L^{k+m})_-\Psi}{\Psi} \\
  &  &    \\
  &=&  - \frac{1}{2} \frac{(ML^{k+1} + \frac{\alpha}{2} L^{k+1} +
\frac{\beta}{2} L^{k+3} )_-{\Psi} }{\Psi} \\
  &  & \\
  &=&  \frac{1}{4} (e^{-\eta} - 1) \frac{(W^{(2)}_k + \alpha
W^{(1)}_{k+1} + \beta W^{(1)}_{k+3} ) \tau }{\tau}
\end{eqnarray*}
and thus (89) holds, ending the proof of Lemma 9.2.

\bigbreak

\noindent \underline{Proof of Theorem 0.2} :
Consider now the shift
$$
s_1 = t_1 + \frac{\alpha}{2} ,~ 3 s_3 = 3t_3 + \frac{\beta}{2},~ \mbox{ all
otherÊ}~ s_i = t_i ,
$$
and define
$$\bar \tau (s) = \tau (t).$$
For even $k \geq -m$, we have
\be
W^{(2)}_k + \alpha W^{(1)}_{k+1} + \beta W^{(1)}_{k+3} = \bar W^{(2)}_k -
\frac{\alpha^2Ê}{4} \delta_{k,-2 },
\ee
where $\bar W_k $ is $W_k $, with $t$ replaced by $s$.  Note the term
$\frac{\alpha^2 }{4} \delta_{k,-2}$ comes from
$$W^{(2)}_{-2} = s^2_1 + \ldots = (t_1 + \alpha/2)^2 + \ldots = W^{(2)}_{-2} +
\alpha W^{(1)}_{-1} + \frac{\alpha^2 }{4} + \ldots.$$
Therefore the equations (89) in Lemma 9.2 get converted into
\be
\bar W^{(2)}_k \bar \tau = \bar c_k \bar \tau, \mbox{ for even } k \geq -m,
\ee
with different constants $\bar c_k $.

Let $\bar \Psi (x,s,z)$ and $\bar \Psi^* (x,s,z)$ be the wave functions
defined by the $\tau$-function $\bar \tau (s)$; note for KdV, we have $
\bar \Psi^* (x,s,z) = \bar \Psi (x,s,-z)$.
Using this fact, combined with the requirement $a_+ b_+ = a_- b_- $,
implies the proportionality of the functions
\begin{eqnarray*}
\sum_{\omega \in \{+,-\}} a_{\omega} \bar \Psi^* (x,s,\omega y) &=&
\sum_{\omega \in \{+,-\}} a_{\omega} \bar \Psi (x,s,-\omega y),\\
\sum_{\omega' \in \{+,-\}} b_{\omega'} \bar \Psi (x,s,\omega' y) &=&
\sum_{\omega' \in \{+,-\}} b_{\omega'} \bar \Psi (x,s,\omega' y) =: \Phi
(x,s,y).
\end{eqnarray*}
The kernel
$$\bar K^E_{x,s} (y,z) := \frac{I_E (z)}{2 y^{1/4} z^{1/4} } \int^x dx~ \Phi
(x,s,y^{1/2} ) \Phi (x,s,z^{1/2} )$$
flows off the kernel (15) at  $s = (\frac{\alpha}{2},
0,\frac{\beta}{6},0,0,\ldots)$.  According to Theorem 0.1, using the Virasoro
equations (91) for $\bar \tau$, the expression
\be
\bar \tau^E := \bar \tau \det(I - \lambda \bar K^E_{x,s} )=e^{-\lb \int_E dz
~Y(s,z,z)}\bar \tau(s)
\ee
 satisfies the equation
$$
\left(-2\sum_{i=1}^{2r}A_i^{\frac{k}{2}+1}\frac{\pl}{\pl
A_i}+\frac{1}{2}(\bar W_k^{(2)}-\bar c_k)\right)\bar\tau^E=0,
\mbox{ for even } k\geq -m,
$$
and, upon undoing the shift above, $\tau^E:=\tau\det(I-\lb
K^E_{x,t})$ satisfies
\be
\left(-2\sum_{i=1}^{2r}A_i^{\frac{k}{2}+1}\frac{\pl}{\pl
A_i}+\frac{1}{2}(W_k^{(2)}+\al W_{k+1}^{(1)}+\beta
W^{(1)}_{k+3}-c_k)\right)\tau^E=0,
\ee
\hfill for even $k \geq -m$,

\medbreak\noindent with principal symbol
$$
\begin{array}{llll}
\sg(W_k^{(2)}+\al W_{k+1}^{(1)}+\beta W_{k+3}^{(1)}-c_k)&=\beta W_{k+3}^{(1)}&
\mbox{if\,\,}\beta\neq 0&k\mbox{\,\,even\,\,}\geq -2\\
&=\alpha W_{k+1}^{(1)}&
\mbox{if\,\,}\beta =0&k\mbox{\,\,even\,\,}\geq 0.
\end{array}
$$

We now apply Lemma 8.1 to the case where all $b_{kk}=0$. Since we are in the
KdV situation, we only need $J_{k_{\ell}}^{(1)}...J_{k_1}^{(1)}$ for odd
$k_i$; this is obtained upon considering the constraints $Q_k\tau^E={\cal
A}_k\tau^E$ for odd $k\geq 1$ and the KdV equations (73) for odd $n\geq
3$. So Lemma 8.1 is applicable and $\tau^E$ satisfies the hierarchy (68) of
KP equations for odd $n\geq 3$, with the commutative $t$-partials replaced by
the non-commutative ${\cal A}_k $-operators.

Note these equations do not have constant terms. Indeed, shifting the
$s$-variables back to the $t$-variables in (92) leads to the appearance
of exponential factors in (92); differentiating that new expression (92)
with regard to the boundary points $A_i$ and letting all $A_i$ tend to
$\iy$ implies that in the limit all terms in (68) tend to zero.

Finally, because of the fact that in
$\tau^E=\tau\det(I-\lb K^E_{0,x})$ the factor $\tau(t)$ is independent of the
boundary points ${\cal A}_i$ and that the equations are free of constant terms,
the Fredholm determinant itself is a solution of equations (16). This ends the
proof of Theorem 0.2.

\section{Examples}

\subsection*{Example 1  (Airy kernel)}

Given an entire function $U(y)$ growing sufficiently fast at $\iy$, consider its
``Fourier" transform,
\begin{equation}
F(u)=\int_{-\iy}^\iy e^{-U(y)+yu}dy~;
\end{equation}
define the associated function
$\rho(z)$ and the differential operator $A:=A_z$:
\be
\rho(z)=\frac{1}{\sqrt{2\pi}}e^{U(z)-zU'(z)}\sqrt{U''(z)}\mbox{ and
} A_z:=\rho(z) \frac{1}{U''(z)}\frac{\pl}{\pl z}  \rho(z)^{-1}.
\ee

\proclaim Proposition 10.1. The function
\be
\Psi(x,z):= \rho(z) F(x+U'(z))
\ee
satisfies the following three equations:
\be
A_z\Psi(x,z)=\frac{\pl}{\pl x}\Psi(x,z),
\ee
\be
\left(U'(\frac{\pl}{\pl x})-x\right)\Psi(x,z) = U'(z) \Psi(x,z).
\ee
and
\be
U'(A)\Psi(0,z)=U'(z)\Psi(0,z),
\ee
with the following asymptotics
\be
\Psi(0,z)= 1+O(z^{-1}) \mbox{ for }~~z \nearrow \iy.
\ee

\noindent \underline{Proof}: At first note that
$$
\frac{\pl}{\pl x}\Psi(x,z)=\rho(z) \frac{\pl}{\pl x}F(x+U'(z))=\rho(z)
\frac{1}{U''(z)}\frac{\pl}{\pl z}F(x+U'(z))=B\Psi(x,z)
$$
Integrating $\frac{\pl}{\pl
y}e^{-U(y)+yu}$ over $\BR$ shows that $F(u)$ satisfies the following
differential equation
\begin{equation}
        U'\left(\frac{\pl}{\pl u}\right)F(u)=uF(u).
        \label{3.2}
\end{equation}
Setting $u=x+U'(z)$ in this equation and multiplying with $\rho(z)$ yield:
$$
\left(U'\left(\frac{\pl}{\pl x}\right)-x\right)\Psi(x,z) = U'(z) \Psi(x,z)
$$
The method of stationary phase applied to
the integral (94) and its derivatives leads to the following estimate,
upon Taylor expanding $U(x)$ around $x=z$, and upon using (95):
$$
      \left(\frac{\pl}{\pl u}\right)^n F(u)\biggr|_{u=U'(z)}\hspace{9cm}
 $$
    \begin{eqnarray}
 &=&\int_{-\iy}^\iy y^ne^{-U(y)+yU'(z)}dy
        \nonumber\\
        &=&\int_{-\iy}^\iy y^n
        e^{-(U(z)+(y-z)U'(z)+\frac{1}{2}(y-z)^2U''(z)+O(y-z)^3)+yU'(z)}dy
        \nonumber\\
        &=&e^{-U(z)+zU'(z)}\int_{-\iy}^\iy y^ne^{-\frac{1}{2}(y-z)^2U''(z)
        \left(1+\frac{U'''}{U''}O(y-z)\right)}dy
        \nonumber\\
&=&e^{-U(z)+zU'(z)}
\frac{1}{\sqrt{U''}}\left(
       \int_{-\iy}^\iy\bigl(\frac{y}{\sqrt{U''}}+z\bigr)^n
        e^{-\frac{1}{2}y^2}dy+O(1/z)\right)
        \nonumber\\
        &=&\rho(z)^{-1}z^n(1+O(1/z)),
        \label{3.4}
\end{eqnarray}
thus ending the proof of proposition 10.1.

\bigbreak

\proclaim Theorem 10.2. For $U(z)=z^3/3$, the function $\Psi(x,z)$, defined by
(96), satisfies the conditions of Theorem 0.2 and the Fredholm determinant
$f(A_1,...,A_{2r})\\ :=\det(I-\lb K_x^E)$ satisfies the  hierarchy of bilinear
partial differential equations
in the $A_i$ for odd $n \geq 3$:
\be
f \cdot \AR_n \AR_1 f-\AR_n f \cdot \AR_1 f-\sum_{i+j=n+1}p_i
(\tilde\AR)f \cdot p_j(-\tilde\AR)f=0,
\ee
with
\be
\AR_n=\sum_{i=1}^{2r}A_i^{\frac{n-1}{2}}\frac{\pl}{\pl A_i},
\quad n=1,3,5,... ;
\ee
 the variables appearing in the Schur polynomials $p_i$ are non-commutative
and are written according to a definite order.
Finally, the first equation in the hierarchy (16) takes on the
following form :
\be
\left({\cal A}_1^3 - 4 ({\cal A}_3 - \frac{1}{2})\right) R +
6({\cal A}_1 R)^2 = 0,
\ee
for
$$
R: = {\cal A}_1 \log f = \sum^{2r}_1 \frac{\partial
\log \det (I - \lambda K^E)}{\partial A_i},
$$
When $E = (- \infty, A),$ the function $R={\cal A}_1 \log f=\frac{\pl }{\pl
A}\log \det (I-K^E)$ satisfies
$$
R^{'''} - 4 AR^{\prime} + 2R + 6{R^{\prime}}^2 = 0 \quad \mbox{\bf
(Painlev\'{e} II)}
$$

\noindent \underline{Proof}: The Painlev\'e II equation for the
logarithmic derivative has been obtained previously by Tracy and Widom
\cite{TW1}; equation (105), which leads to Painlev\'e, is
new. For
$U(z)=\frac{1}{3}z^3$, we have
$~\al=0,~\beta=4,~m=2$, and $V(z)=\frac{2}{3}z^3$; also
\be
A:=\frac{1}{2z}\left( \frac{\pl}{\pl
z}+2z^2\right)-\frac{1}{4}z^{-2}
\quad \mbox{and} \quad \hat A=\frac{\pl}{\pl x}.
\ee
Then in terms of the Airy function
$$F(u) := \int_{-\infty}^{\infty} e^{-\frac{y^3}{3} + yuÊ} dy,$$
$\Psi (x,z)$ has, according to proposition 10.1, the following expression
$$\Psi (x,z) = \frac{1}{\sqrt{\pi}} e^{-\frac{2}{3} z^3} \sqrt{z} F (x + z^2 )$$
and is a solution of
$$
A \Psi (x,z) = \hat A \Psi (x,z),
$$
with $\Psi (0,z)$ satisfying
$$
B^2 \Psi (0,z) = z^2 \Psi (0,z) \mbox{ and } \Psi (0,z) = 1 + O (z^{-1} ).
$$
Setting $b_+ = 1$ and $b_- = 0$, we find for $\Phi (x,u)$ and $K^E_x (y,z)$ in
(15),
$$
\Phi (x,u) = \frac{\sqrt{u}}{\sqrt{\pi}} A (x + u^2)
$$
$$
K^E_x (y,z) = I_E (z) \frac{1}{2 \pi} \int^x A (x + y) A (x + z) dx.
$$
Then $f(A_1 ,\ldots,A_{2r} ) := \det (I - \lambda K^E_x )$ satisfies the
hierarchy of equations (16), with lower weight terms; the ${\cal A}_n $ are
defined by (107).  However, upon rewriting the variables of $p_n $ in an
appropriate way, all lower weight terms can be removed, as follows from a
combinatorial argument.

Finally $\tau^E:=\tau \det(I-\lb K^E_{x,t})$ satisfies equation (93) for
$\al=0, ~\beta=4$ and $m=2$; in particular for $k=2$ and $0$, one finds, upon
dividing by $\tau$:

$$
{\cal A}_1 \log \tau^E=
\frac{1}{2} \left(\sum_{i\geq 3} it_i \frac{\pl}{\pl t_{i-2}} +
2\frac{\pl}{\pl t_1} \right)\log \tau^E +
\frac{t_1^2}{4}
$$
$$
{\cal A}_3 \log \tau^E=
\frac{1}{2} \left(\sum_{i\geq 1} it_i \frac{\pl}{\pl t_{i}}+2\frac{\pl}{\pl t_3}
  \right)\log \tau^E + \frac{1}{16}
$$
from which the partial derivatives $\frac{\partial}{\partial t_1} \log \tau^J, \frac{\partial}{\partial
t_3}  \log \tau^J$ and $\frac{\partial^2}{\partial t_1 \partial t_3}\, \log
\tau^J \, \mbox{at}\, t = 0$  can be extracted. Putting these partials in the
KP-equation (73) leads to the equation (105). In the particular case of a
semi-infinite interval, one finds the Painlev\'e II equation, as stated in
Theorem 10.1.

\subsection*{Example 2 (Bessel Kernel)}

In the next theorem, we investigate the PDE's for the Fredholm
determinant of the Bessel kernel, depending on a parameter $\nu$;
special values of $\nu$ yield the sine kernel, which was originally
investigated by Jimbo, Miwa, Mori and Sato \cite{JMMS}. The equation (109)
is new, although in the case of an interval the Painlev\'e equations where
obtained in \cite{JMMS} and \cite{TW1}.

\bigbreak

\proclaim Theorem 10.3. Given the kernel
\be
K^E(y,z)=-{1\over2}I_E(z) \int_0^1sJ_\nu(s\sqrt{y})
J_\nu(s\sqrt{z})ds,
\ee
the Fredholm determinant $f(A_1,...,A_{2r}):=\det(I-\lb K_x^E)$ satisfies the
hierarchy (16) of bilinear partial differential equations
in the $A_i$ for odd $n \geq 3$, with
\be
\AR_n:=\sum_{i=1}^{2r}A_i^{\frac{n+1}{2}}\frac{\pl}{\pl A_i},
\quad n=1,3,5,... ;
\ee
The first equation in the hierarchy (16) for $F:=\log \det (I-\lb K^E)$ takes
on the following form:
\be
\left({\cal A}_1^4 - 2 {\cal A}_1^3 + ( 1 - \nu^2) {\cal A}_1^2 + {\cal A}_3
\left({\cal A}_1 - \frac{1}{2}\right)\right) F - 4({\cal A}_1 F) ({\cal A}_1^2
F) +  6({\cal A}_1^2 F)^2 = 0
\ee
when $E = (0, A),$ we have for $R: = -{\cal A}_1F=-A~ \frac{\pl}{\pl A}\log
\det (I-\lb K^E),$ the equation:
$$
 A^2 R^{'''} + AR^{''} + (A - \nu^2) R^{'} - \frac{R}{2} + 4RR^{'} - 6
AR^{'2} = 0\,  \mbox{  \bf (Painlev\'e V)}
$$ \par

\medbreak

\noindent \underline{Proof} :
Pick $V(z) = -z$; then $m = 0$, $\alpha = - 2$, $\beta = 0$ and
$$A_z = \frac{1}{2} z (\frac{\partial}{\partial z} - 1) \quad \mbox{ and }
\quad \hat A_x = \frac{1}{2} (x - 1) \frac{\partial}{\partial x}$$
We look for a function $\Psi (x,z)$ satisfying
\be
A_z \Psi (x,z) = \hat A_x \Psi (x,z)
\ee
with initial condition $\Psi (0,z)$ satisfying
\be
(4 A^2_z - 2A_z - \nu^2 + \frac{1}{4}) \Psi (0,z) = z^2 \Psi (0,z),Ê\quad \Psi
(0,z) = 1 + O (\frac{1}{z})
\ee
The solution to the differential equation (111) is given by
\begin{eqnarray*}
\Psi (0,z) = B(z) &=& \varepsilon \sqrt{z} H_{\nu} (iz) \\
     &  & \\
     &=& \frac{e^z 2^{\nu + 1/2}}{\Gamma (-\nu + 1/2)} \int_1^{\infty}
\frac{z^{-\nu + 1/2} e^{-uz}}{(u^2 - 1)^{\nu+1/2}} du \\
     &  & \\
     &=& 1 + O (\frac{1}{z})
\end{eqnarray*}
with $\varepsilon = i \sqrt{\pi/2} e^{i \pi \frac{\nu}{2} }$ , $- \frac{1}{2} <
\nu < \frac{1}{2}$.  Then
$$\Psi (x,z) = e^{xz} B ((1 - x) z)$$
satisfies (110); from Theorem 0.2, $\Psi (x,z)$ satisfies a second
order spectral problem, which can explicitly be computed :
$$
\left(\frac{d^2 }{dx^2 } - \frac{(\nu^2 - \frac{1}{4})}{(x - 1)^2
}\right) \Psi (x,z) = z^2 \Psi (x,z).
$$
Picking $b_+ = e^{- i \pi \nu/2}/2 \sqrt{\pi}$
and
$b_- = i \bar b_+$,
 yield for (15):
\begin{eqnarray*}
\Phi (x,z) &=& \frac{e^{-i \pi \nu/2 }}{2 \sqrt{\pi}} e^{-z} \Psi
(x,z) + \frac{ie^{i \pi \nu/2}}{2 \sqrt{\pi}} e^z \Psi (x,-z)\\
   &= &\sqrt{\frac{(x-1)z}{2}} J_{\nu} ((1 - x)iz),
\end{eqnarray*}
and
\begin{eqnarray*}
K^E_x &=& I_E (z) \int_1^x \frac{\Phi (x,\sqrt{y}) \Phi
(x,\sqrt{z})}{2 y^{1/4} z^{1/4} } dx\\
  &=& -\frac{1}{2} I_E (z) \int_0^{(1 - x)i} s J_{\nu} (s \sqrt{y})
J_{\nu} (s \sqrt{z}) ds.
\end{eqnarray*}
The special value $x = i + 1$ leads to the standard Bessel kernel:
\begin{eqnarray*}
K^E_{1 + i} &= &- \frac{1}{2} I_E (z) \int_0^1 s J_{\nu} (s \sqrt{y})
J_{\nu} (s \sqrt{z}) ds\\
 &=& I_E (z) \frac{J_{\nu} (\sqrt{y}) \sqrt{z} J'_{\nu} (\sqrt{z}) -
J_{\nu} (\sqrt{z}) \sqrt{y} J'_{\nu} (\sqrt{y})} {2(z - y)}.
\end{eqnarray*}
From Theorem 0.2, the Fredholm determinant
$$
f(A_1, \ldots ,A_{2r}) := \det (I - \lambda K^E_{1+i})
$$
satisfies equation (16), with
$$
{\cal A}_n = \sum_{i=1}^{2 \tau} A^{\frac{n+1}{2}}_{i} \frac{\pl}{\pl
A_i }, ~~~n = 1,3,5,\ldots.
$$
Picking the special value $x = i \in 1$, leads to the shift $t_1 \mapsto t_1
+ i + 1$; setting $\alpha = - 2$ and $\beta = 0$ in (93) and shifting
$t_1 $, as above, leads to
$$
{\cal A}_1 \log \tau^E=
\frac{1}{2} \left(\sum_{i\geq 1} it_i \frac{\pl}{\pl t_{i}} +
\sqrt{-1}\frac{\pl}{\pl t_1} \right)\log \tau^E +
\frac{1}{4}(\frac{1}{4}-\nu^2)
$$
$$
{\cal A}_3 \log \tau^E=
\frac{1}{2} \left(\sum_{i\geq 1} it_i \frac{\pl}{\pl t_{i+2}}+
\frac{1}{2}\frac{\pl^2}{\pl t_1^2}+\sqrt{-1}\frac{\pl}{\pl t_3}   \right)
\log \tau^E + \frac{1}{4}\left(\frac{\pl}{\pl t_1}\log
\tau^E\right)^2;
$$
expressing the partial derivatives appearing in the first KP-equation for $p=2$
at $t=0$ in terms of the operators ${\cal A}_1=\sum_1^{2r}A_i\frac{\pl}{\pl
A_i}$ and ${\cal A}_3=\sum_1^{2r}A_i^2\frac{\pl}{\pl A_i}$ leads to the partial
differential equation (109), which for $E = (0,A)$ leads to the
Painlev\'e V equation, ending the proof of Theorem 10.3.

\bigbreak

\noindent \underline{Sine kernel} :
Note that for $\nu = \pm 1/2$,
\begin{eqnarray*}
\frac{2}{\pi} \frac{k^- (y,z)}{\sqrt{yz}} &=& K^{\pm 1/2}_{x,0} (y^2
,z^2 )\\
   &= &    \frac{1}{\pi} \int_0^x
\frac{\renewcommand{\arraystretch}{0,5}
\begin{array}[t]{c}
\sin\\
\cos
\end{array}
\renewcommand {\arraystretch}{1}
 xy
\renewcommand{\arraystretch}{0,5}
\begin{array}[t]{c}
\sin\\
\cos
\end{array}
\renewcommand {\arraystretch}{1} xz}{y^{1/2} z^{1/2}} dx \\
   &=&Ê\frac{1}{4 \pi} \int_0^x (e^{-ixy} \mp e^{ixy} ) (e^{ixz} \mp
e^{-ixz} ) dxÊ\\
   &= &\frac{1}{2 \pi} (\frac{\sin x(y - z)}{y - z} \mp \frac{\sin x(y
+ z)}{y + z})
\end{eqnarray*}
Note this kernel can be obtained in a direct way, as follows: assume $\tau=1$,
$p=2$, $a_1=b_1=1$ and $a_{-1}=b_{-1}=\pm1$ in (5);
one finds
\begin{eqnarray*}
k^\pm_t(y,z)&:=&k_t(y,z)\\
&=&\int_0^x\left(\Psi^*(ix,t,y)\pm\Psi^*(ix,t,-y)\right)
\left(\Psi(ix,t,z)\pm\Psi(ix,t,-z)\right) dx
\end{eqnarray*}
where
$$
\Psi(ix,t,z)=e^{ixz}e^{\sum^{\infty}_{1} t_k z^k}\quad\mbox{and}\quad
\Psi^*(ix,t,z)=e^{-ixz}e^{-\sum^{\infty}_{1} t_k z^k}.
$$
Note the kernel $k_t^\pm(y,z)$ flows off the kernel at $t=0$:
\begin{eqnarray*}
k^\pm_{x,0}(y,z)&=&\int^x_0 \left(e^{-ixy}\pm e^{ixy}\right)
 \left(e^{ixz}\pm e^{-ixz}\right)dx\\
&=&2 \left( \frac{\sin x(y-z) }{y-z}
\pm \frac{\sin x(y+z) }{y+z} \right).
\end{eqnarray*}
Moreover
$$
\det\left(I-\lb k_{x,t}^{[-a,a]}\right)=\det\left(I-2\lb
k_{x,t}^{[0,a]}\right).
$$
It satisfies
\bea
\left(-a^{2k+1} \frac{\pl}{\pl a}+{1\over2}J_{2k}^{(2)}\right)
\det\left(I-\lb k_t^{[-a,a]}\right)
=0\quad\hbox{for}\quad k\ge 0,\label{70}
\eea
since $J_{2k}^{(2)}1=0$ for $k\geq 0$.  Also for
$K(\lb,\lb')=(1/2)k(z,z')/\sqrt{zz'}$, $\lb=z^2$, $\lb'=z'^2$,
we have
$$
\left(-A^{k+1}{\pl\over\pl A}+{1\over4}J_{2k}^{(2)}\right)
\det(I-2\lb K_{x,t}^{[0,A]})=0.\label{71}
$$

\end{document}